\documentclass[11pt]{article}
\pdfoutput=1 % if your are submitting a pdflatex (i.e. if you have  images in pdf, png or jpg format)
\usepackage{amssymb, amsmath,amsfonts}
\usepackage{graphicx}
\usepackage{xcolor}
\usepackage{color}
\usepackage[pdftex, bookmarks=true,colorlinks,linkcolor=red,urlcolor=blue,citecolor=blue]{hyperref}
\usepackage[normalem]{ulem}
\usepackage{cite}
\usepackage{subcaption}
\usepackage{slashed}

\def\arXiv#1{\href{http://arxiv.org/abs/#1}{arXiv:#1}}
\def\arXiv#1#2{\href{http://arxiv.org/abs/#1}{arXiv:#1}}

\def\doi#1#2{\href{http://doi.org/#1}{#2}}

\allowdisplaybreaks
\usepackage{enumerate}
\usepackage[normalem]{ulem}

\makeatletter\@addtoreset{equation}{section}\makeatother

\newcommand{\preprint}[1]{\begin{table}[t]  %%
             \begin{flushright}               %%
             {#1}                             %%
             \end{flushright}                 %%
             \end{table}}                     %%
\renewcommand{\title}[1]{\vbox{\center\LARGE{#1}}\vspace{5mm}}
\renewcommand{\author}[1]{\vbox{\center#1}\vspace{5mm}}

\newcommand{\address}[1]{\vbox{\center\em#1}}

\usepackage{bm}
\def\mI{{\bf I}}

\def\be{\begin{eqnarray}}
\def\ee{\end{eqnarray}}
\def\bea{\begin{eqnarray}}
\def\eea{\end{eqnarray}}

\def\Dslash{\,\,{\raise.15ex\hbox{/}\mkern-12mu D}}
\def\Dbarslash{\,\,{\raise.15ex\hbox{/}\mkern-12mu {\bar D}}}
\def\delslash{\,\,{\raise.15ex\hbox{/}\mkern-9mu \partial}}
\def\delbarslash{\,\,{\raise.15ex\hbox{/}\mkern-9mu {\bar\partial}}}
\def\pslash{\,\,{\raise.15ex\hbox{/}\mkern-9mu p}}
\def\calDslash{\,\,{\raise.15ex\hbox{/}\mkern-12mu {\cal D}}}

\def\lae{\mathrel{\mathop{\smash{\lower .5 ex \hbox{$\stackrel<\sim$}}}}}
\def\lae{\mathrel{\mathop{\smash{\lower .5 ex \hbox{$\stackrel>\sim$}}}}}

\begin{document}

\unitlength = .8mm

\begin{titlepage}
\vspace{.5cm}
\preprint{}
%{\color{violet}{\large{\today}}}
\begin{center}
\hfill \\
\hfill \\
\vskip 1cm

\title{\bf Coexistence of topological semimetal states in holography}
\vskip 0.5cm

{Haoqi Chu$^{a,b}$}, {Xuanting Ji$^{b,c}$}\footnote{Email: {\tt jixuanting@cau.edu.cn}}, {Ya-Wen Sun$^{c,d}$}\footnote{Email: {\tt yawen.sun@ucas.ac.cn}}

\address{${}^a$ College of Land Science and Technology, \\
China Agricultural University, Beijing 100193, China}
\vspace{-10pt}
\address{${}^b$Department of Applied Physics, College of Science, \\
China Agricultural University, Beijing 100083, China}
\vspace{-10pt}
\address{${}^c$School of Physical Sciences, and CAS Center for Excellence in Topological Quantum Computation, University of Chinese Academy of Sciences, %Zhongguancun east road 80,
Beijing 100049, China}
\vspace{-10pt}
\address{${}^d$Kavli Institute for Theoretical Sciences, \\
University of Chinese Academy of Sciences, %Zhongguancun east road 80,
Beijing 100049, China }
\vspace{-10pt}

\end{center}
%\vskip 1.5cm
\vspace{-10pt}
\abstract{We introduce a holographic model that exhibits a coexistence state of the Weyl semimetal and the topological nodal line state, providing us with a valuable tool to investigate the system's behavior in the strong coupling regime. Nine types of bulk solutions exhibiting different IR behaviors have been identified,  corresponding to nine different types of boundary states. These nine states include four distinct phases, namely the Weyl-nodal phase, the gap-nodal phase, the Weyl gap phase and the gap-gap phase, four phase boundaries, which are the Weyl-Dirac phase, the gap-Dirac phase, the Dirac-gap phase and the Dirac-nodal phase, and finally a double critical point. A phase diagram 
is plotted that exhibits qualitative similarity to the one obtained in the weak coupling limit. The anomalous Hall conductivity, which serves as an order parameter, and the free energy are calculated, with the latter showing the continuity of the topological phase transitions within the system. Our study highlights the similarities and differences in such a topological system between the weak and strong coupling regimes, paving the way for further experimental observations.}

\vfill

\end{titlepage}

\begingroup
\hypersetup{linkcolor=black}
\tableofcontents
\endgroup
%%%%%%%%%%%%%%%%%%%%%%%%%%
\section{Introduction}
\label{sec:1}
%%%%%%%%%%%%%%%%%%%%%%%%%%
Classifying the states of many-body systems is one of the central problems in condensed matter physics. The topological properties of many-body systems have become increasingly important in recent years, giving rise to a new class of many-body systems known as topological materials. In contrast to traditional materials, the phase transition in a many-body system with non-trivial topological properties cannot be described by the Landau paradigm. This is an indication of the existence of new physics and attracts a lot of research interest in the study of the classification of the many-body system from the topology\cite{zhida}, as well as the possible topological phase transition process.   

Topological semimetals are a large class of topological materials with several novel properties, such as non-dissipative transport, topological robustness, and the realization of particles that cannot be present in the standard model of particle physics. They are characterized by the crossing of the conduction and valence bands in the Brillouin zone. In addition, various symmetries play important roles in different types of semimetals. For instance, Dirac semimetals possess both time reversal and inversion symmetries\cite{zkliu,Ylchen,kane}, while Weyl semimetals possess only time reversal or inversion symmetry\cite{Wan,rmb}. Topological nodal line semimetals are characterized by mirror reflection symmetry, PT symmetry, or glide mirror symmetry\cite{burkov1,fang1}, and Weyl-$Z_2$ semimetals have an additional $Z_2$ symmetry\cite{Gorgar1,Gorgar2}.
The lattice model based on the symmetry can be used to describe the fundamental properties of topological systems. However, it is only valid to describe the low energy excitation of quasiparticles, in particular in the regime of weak coupling. In realistic materials, inevitably strong interactions such as spin-orbit coupling and perturbations exist, which invalidate the quasiparticle description. Besides, hydrodynamic behaviours of the Weyl semimetals have been observed in the experiments\cite{Gooth}. For the study of physical problems, especially those related to physical systems in the strong coupling regime, a distinct system description is necessary.

Holography, originated from string theory, is an efficient way to study strongly coupled many body systems\cite{Hartnoll:2016apf,Landsteiner:2019kxb}. Holographic models have been successfully used to describe superconductors\cite{Hartnoll:2008kx}, Dirac semimetals\cite{Grignani:2016npu}, Weyl semimetals\cite{Landsteiner:2015lsa,Landsteiner:2015pdh}, nodalline semimetals\cite{Liu:2018bye,Liu:2020ymx}, $Z_2$-Weyl semimetals\cite{Ji:2021aan}, and others. Besides, transports\cite{Landsteiner:2011cp,Landsteiner:2011iq,Ji:2019pxx,Gao:2023zbd} and the effect of disorders\cite{Park:2022mxj,Huang:2023ihu,Wang:2023rca,Ahn:2024ozz} also can be studied by holography. Holography builds the bridge that connects high-energy physics and the condensed matter world. It allows for the study of similarities and differences between weak coupling and strong coupling regimes, promoting a deeper understanding of topological systems and gravity.

TaAs is considered the first realized Weyl semimetal candidate. It can present the Weyl semimetal state and the nodal line state, depending on the existence of spin-orbit coupling\cite{weng}. Besides, from first principle calculations, it has been found that a collinear antiferromagnet can host both Weyl semimetal and nodal line semimetal\cite{zhan}. Therefore, it would be interesting to verify if we could find a strongly coupled system that can host both the Weyl semimetal and the nodal line semimetal states.

To construct such a holography model, we will need to obtain clues from a covariant formulation of the system in the weakly coupled limit, which could be fulfilled by Lorentz-covariant field theories. Lorentz-covariant field theory models have been successfully applied in the holography model of the Weyl semimetal\cite{Landsteiner:2015lsa,Landsteiner:2015pdh} and the nodal line semimetals \cite{Liu:2018bye,Liu:2020ymx}. Recently, a field theoretic model based on Lorentz covariant field theory demonstrated the coexistence of a Weyl semi-metal phase and a topological nodal line semi-metal\cite{Ji:2023rua}. {The model introduces an eight-component spinor that exists in an expanded Hilbert space. Each half of the eight-component spinor is responsible for forming either the Weyl nodes or the nodal ring of a nodal line semimetal.} In this paper, we construct a holographic model for studying a physical system in which the Weyl semimetal and the nodal-line semimetal coexist in the strongly coupled regime. By solving the model at zero temperature we obtain nine distinct solutions corresponding to nine different phases, similar to the weakly coupled field theory models. The anomalous Hall conductivity is calculated as one order parameter of the topological phase transition. Additionally, we assessed the continuity of the phase transition by calculating the free energy. 

This paper is organized as follows: Section \ref{sec:2} reviews effective field theory models that realize the coexistence of Weyl semimetals and nodal line semimetals. Section \ref{sec:31} presents the holographic model and its solutions at zero temperature, while Section \ref{sec:32} provides the complete phase diagram. Anomalous Hall conductivity and free energy calculations are presented in Section \ref{sec:33}. Finally, Section \ref{sec:4} offers discussions and outlook. 
%%%%%%%%%%%%%%%%%%%%%%%%%%
\section{Review of effective field theory model of coexistence of topological semimetal states}
\label{sec:2}

For our goal of developing a holographic model for topological semimetals with the coexistence of topological nodal line and Weyl semimetals, we will need to initially examine an effective field theoretic model in a covariant form, which provides insights for the holographic construction of the system. Furthermore, it is essential to compare the physics of strongly coupled systems with that of weakly coupled systems to gain a comprehensive understanding of the strongly coupled effects in topological states. In this section, we therefore provide a review for the effective field theory model in which a topological system where Weyl semimetals and nodal line semimetals can coexist\cite{Ji:2023rua}. We will present a description of the phase diagram along with discussions of phase transitions to provide a comprehensive understanding of the behavior of the weakly coupled system.

\subsection{Model and Energy Spectrum}
\label{sec:21}

To have a field theoretic model of topological semimetals in a covariant form, Lorentz symmetry breaking terms in a relativistic field theory are required. 
The Lorentz breaking relativistic field theory models have been studied extensively in high energy physics. {The energy bands of several topological semimetal states can be realized by adding Lorentz symmetry breaking terms, which can be viewed as the effective mass terms of the fermions in the relativistic field theory.} For instance, a Weyl semimetal with a single pair of Weyl nodes can be achieved by introducing a Lorentz-breaking term proportional to the gamma matrices $\gamma_{\mu}\gamma_{5}$\cite{Grushin:2012mt,Grushin:2019uuu}. Similarly, a nodal line semimetal can be created by adding a term proportional to an antisymmetric two-form field to break the Lorentz symmetry\cite{Liu:2018bye,Liu:2020ymx,Burkov}.

Therefore, to incorporate both Weyl and nodal line semimetal states in one system, we need a combination of these two kinds of terms, i.e. the axial gauge field responsible for the Weyl semimetal and a two-form field of the nodal line semimetal. In \cite{Ji:2023rua}, the competition between these two terms in an effective field theory has been considered. It was shown that the system behaves differently when the non-zero components of the one-form and the two-form fields are in two nonequivalent configurations. Specifically, the coexistence of the nodal line state and the Weyl semimetal state occurs only when the non-zero component of the axial gauge field aligns with the axis perpendicular to the plane where the nodal ring persists. Conversely, only the Weyl semimetal state survives if the non-zero component of the axial gauge field lies within the plane of the nodal ring. This is attributed to the fact that the existence of topological nodal line semimetals heavily relies on the preservation of certain symmetries, the mirror reflection symmetry in this context. This mirror reflection symmetry will be broken by the axial gauge field. The holographic study for this system will be presented in a future work. 

In this paper, we focus on the case of coexistence of both semimetal states when the axial gauge field lies in the plane of the two-form field which breaks the mirror symmetry. In order to solve the problem of destroyed nodal line state and find coexisted semimetal states in this case, an enlarged Hilbert space spanned by an eight component spinor was considered in \cite{Ji:2023rua}. \footnote{Different from the {effective field theory} model raised in\cite{Ji:2023rua}, here we have considered the contributions of a pure imaginary field according to the duality relation raised in \cite{Liu:2020ymx}.}. This introduction of an eight component spinor is similar to the case of the Z$_2$-Weyl semimetal where an extra spin degree of freedom in the fermion was involved \cite{Grushin:2012mt, kimb, Burkov, Colladay:1998fq,Ji:2021aan}. We define the generalized $8\times8$ gamma matrices as
\be
\label{eq:88marix}
\Gamma^{\mu}\equiv\gamma^{\mu}\otimes \mathbb{I}_2\,,~~\Gamma^{5}\equiv\gamma^{5}\otimes \mathbb{I}_2\,,\ee
where $\mu=0,1,2,3$, $\gamma^\mu$ is the $4\times 4$ Dirac Gamma matrix and $\mathbb{I}_2$ is the $2\times2$ unit matrix. With these generalized $8\times8$ matrices and an eight-component spinor $\Psi$ which describes the fermions, a Lagrangian can be obtained which could have Weyl and nodal line semimetal states simultaneously even when the axial gauge field lies within the plane of the nodal ring
\be\label{eq:1Lagrangian}
\mathcal{L}&=&\Psi^{\dagger}\Gamma^{0}\left[\left(i\Gamma^{\mu}\partial_{\mu}+\Gamma^{\mu\nu}b_{\mu\nu}+\Gamma^{\mu\nu}\Gamma^{5}b_{\mu\nu}^{5}+M_1\right){\mI}_{1}+\left(i\Gamma^{\mu}\partial_{\mu}-\Gamma^{0}\Gamma^{\mu}\Gamma^5 b_{\mu} + M_2\right){\mI}_{2}\right]\Psi,
\ee
where $\Gamma^{\mu\nu}=\frac{i}{2}[\Gamma^\mu, \Gamma^\nu]\,$. $ {\mI}_{1}$ and $ {\mI}_{2}$ are two diagonal matrices with diagonal elements as $ {\mI}_1=\text{diag}\left(1,0,1,0,1,0,1,0\right)$ and $ {\mI}_2=\text{diag}\left(0,1,0,1,0,1,0,1\right)$, respectively. $M_1$ and $M_2$ are two mass terms. {$b_{\mu}$ is an axial gauge field responsible for separating two Weyl nodes, and the term $\Gamma^{\mu}\Gamma^5 b_{\mu}$ is a Lorentz breaking term supporting the energy bands of the Weyl semimetal. $b_{\mu\nu}$ is an antisymmetric real two-form field, and the term $\Gamma^{\mu\nu}b_{\mu\nu}$ contributes to the formation of the nodal line semimetal. According to the duality relation between the anti-symmetric tensor $\bar{\psi}\Gamma^{\mu \nu}\Gamma^5\psi=-\frac{i}{2}\varepsilon^{\mu\nu}_{~~\alpha\beta}\bar{\psi}\Gamma^{\alpha \beta}\psi\,$ in the four-dimensional Minkowski spacetime, a pure imaginary dual part of the $b_{\mu\nu}$ needs to be introduced, which is $b_{\mu\nu}^{5}$. When the values of $b_{\mu\nu}$ as well as $b_{\mu\nu}^5$ are zero but $b_{\mu}$ has a nonzero component, the Weyl semimetal spectrum is obtained. While in the case when $b_{\mu}$ is zero and $b_{\mu\nu}$ as well as $b_{\mu\nu}^5$ are nonzero, a nodal ring forms. The coexistence of the Weyl semimetal and the nodal line semimetal can be obtained by turning on both the one-form and the two-form fields.}

{Without loss of generality, we can set the non-zero component of the two-form field as $b_{xy}$, then the non-zero component of the imaginary dual field $b_{\mu\nu}^{5}$ is $b_{tz}^{5}=i b_{xy}$. This choice results in a nodal ring in the $x-y$ plane. On the other hand, we have two choices for the non-zero component of $b_{\mu}$, one along the $z$ direction and the other along the $x$ (or $y$) direction. The Weyl nodes will be on the axis of the direction of the non-zero component of $b_{\mu}$ in this model. To show the coexistence of the Weyl semimetal and the nodal semimetal intuitively, we will set $b_{x}$ to be nonzero.} The energy spectrum of the eight eigenstates in this Lagrangian can be solved at the $k_z=0$ plane with a nonzero constant $b_{xy}$ component of the two-form field and $b_{x}$
\be\label{ospectrum}
E_{1n\pm}=\pm\sqrt{\left(\sqrt{k_{x}^2+k_{y}^2+M_{1}^2}\mp4b_{xy}\right)^2}\,,
E_{2w\pm}=\pm\sqrt{\left(\sqrt{k_{x}^2+M_{2}^2}\mp b_{x}\right)^2+k_{y}^2}\,.
\ee

\subsection{Phase Diagram}
\label{sec:22}

\begin{figure}[ht!]
  \centering
\includegraphics[width=0.8\textwidth]{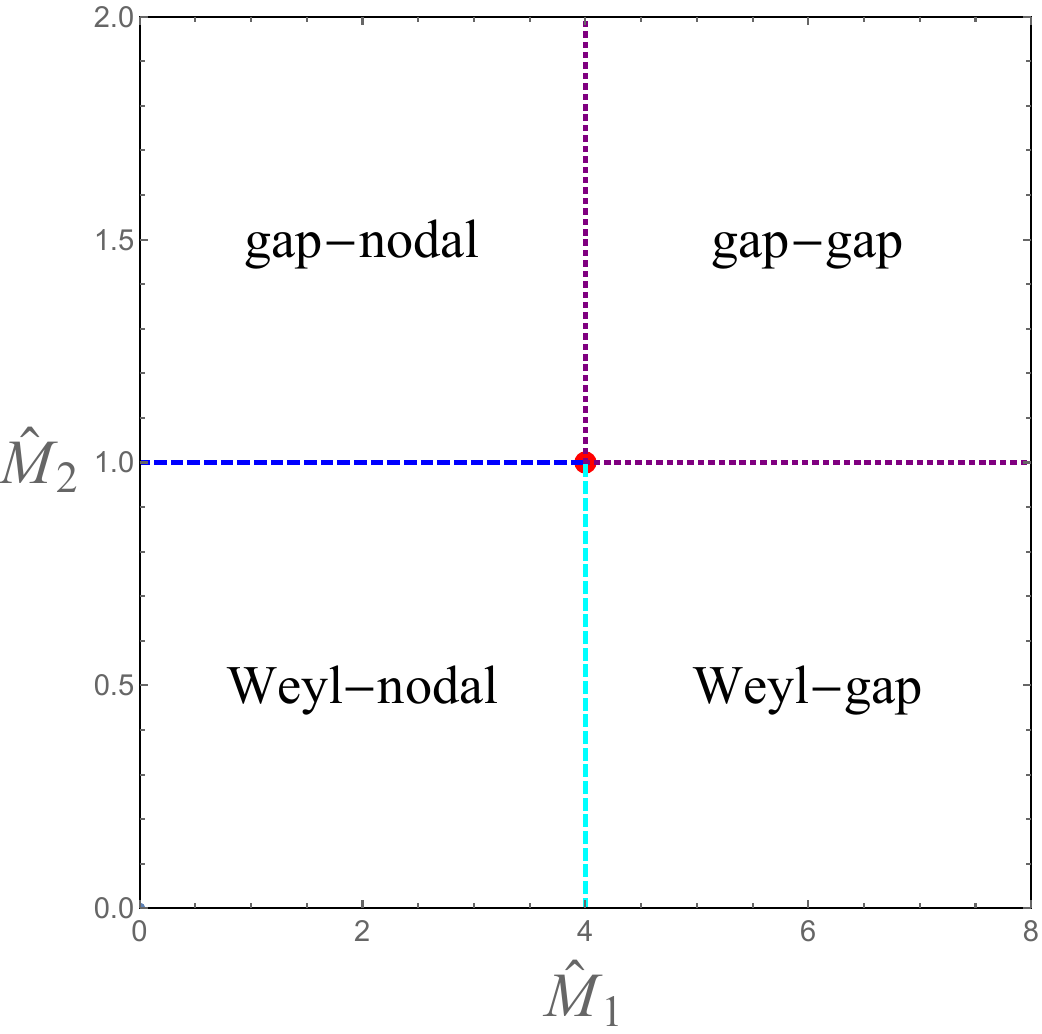}
\vspace{-0.3cm}
  \caption{\small The phase diagram of the system \eqref{eq:1Lagrangian} with three dimensionless parameters $\hat{M}_1=M_1/c$, $\hat{M}_2=M_2/b$ and $c/b=1$. The red point is the double critical point at which both Weyl nodes and nodal ring become critical (Fig. \ref{fig:eight}(d)). The horizontal blue dashed line corresponds to the critical phase in which the Weyl nodes annihilate into a critical Dirac node while a nodal ring still exists (Fig. \ref{fig:eight}(e), the nodal-critical phases). The vertical cyan dashed line corresponds to the critical phase in which the radius of the nodal ring becomes zero while a pair of Weyl nodes still exists (Fig. \ref{fig:eight}(b), the Weyl-critical phases). The purple dotted lines correspond to the phase where the pair of Weyl nodes annihilates into a critical Dirac point(or the radius of the nodal ring becomes zero) while the nodal line (or the pair of Weyl nodes) becomes gapped (Fig. \ref{fig:eight}(f)(or (h)), the critical-gap phase).}
 \label{fig:01}
\end{figure}

Eight eigenstates in Eq.\eqref{ospectrum} could be divided into two groups. $E_{1n\pm}$ is responsible for forming the nodal ring while $E_{2w\pm}$ is responsible for the Weyl nodes. Without loss of generality, the values of the one and the two form field can be set as $b_{x}=b\delta^{\mu}_{x}$ and $b_{xy}=c$. The effective radius of the nodal ring in the nodal line semimetal state can be defined as $\sqrt{16c^2-M_{1}^2}$, while the distance between two Weyl nodes can be obtained as $\sqrt{b^2-M_{2}^2}$. Changing the values of $M_1$, $M_2$, $b$ and $c$, different phases of the system can be found. The corresponding phase diagram is shown in Fig.~\ref{fig:01} with three dimensionless parameters $\hat{M}_1=M_1/c$, $\hat{M}_2=M_2/b$ and $c/b$\footnote{Without loss of generality, we fix $c/b=1$ in Fig.~\ref{fig:01}}. The diagram for the energy spectrum of each phase is presented in Figure \ref{fig:eight} in Appendix \ref{app:a}. 

\vspace{0cm}

In Fig.~\ref{fig:01}, the red point is the critical point at which both the Weyl semimetal and the nodal ring become critical and shrink to a point in Fig.~\ref{fig:eight}(d). The blue dashed line corresponds to the phase transition line where the Weyl nodes annihilate into a critical Dirac node while a nodal ring still exists in Fig.~\ref{fig:eight}(e). The cyan dashed line corresponds to the phase transition line where the radius of the nodal ring becomes zero while a pair of Weyl nodes still exist in Fig.~\ref{fig:eight}(b). The purple dotted lines correspond to another type of phase transition lines where the pair of Weyl nodes annihilates into a critical Dirac point(or the radius of the nodal ring becomes zero) while the nodal line (or the pair of Weyl nodes) becomes gapped in Fig.~\ref{fig:eight}(f)(or (h)). The down-left portion of the phase diagram corresponds to the phase in Fig.~\ref{fig:eight}(a). The up-left portion of the phase diagrams corresponds to the phase in Fig.~\ref{fig:eight}(g). The down-right portion of the phase diagrams corresponds to the phase in Fig.~\ref{fig:eight}(c). The up-right portion of the phase diagram corresponds to the phase Fig.~\ref{fig:eight}(i).

In summary, in the 8-component spinor model Eq.\eqref{eq:1Lagrangian}, the Weyl semimetal and the nodal line semimetal states could coexist and several possible topological phase transitions will happen. In the next section, we will construct a holographic model to investigate such a coexistence state in the strong coupling regime and compare the similarities and differences of the topological phase transitions in the weak and strong coupling regime.

%%%%%%%%%%%%%%%%%%%%%%%%%%

%%%%%%%%%%%%%%%%%%%%%%%%%%
\section{Holographic coexistence of  topological nodal line and Weyl semimetals}
\label{sec:3}
%%%%%%%%%%%%%%%%%%%%%%%%%%
Inspired by the weakly coupled model \eqref{eq:1Lagrangian}, this section presents a holographic model to investigate the coexistence state of the Weyl semimetal and the nodal line semimetal in the strong coupling regime. The model will be given in section \ref{sec:31} and then it will be solved at zero
temperature in section \ref{sec:32}. Nine solutions have been found that correspond to nine different phases, which is the same as in the weakly coupled model. A similar phase diagram is also obtained. Section \ref{sec:33} will give additional details regarding the topological phase transition, including the free energy and the anomalous Hall conductivity. 
\subsection{The holographic set-up}
\label{sec:31}
Based on the dictionary of the gauge/gravity duality, a holographic model to realize a state where the Weyl semimetal and the nodal line semimetal coexist can be obtained as follows
\bea
\begin{aligned}
S&=\int d^5x\sqrt{-g}\bigg[\frac{1}{2\kappa^2}\bigg(R+\frac{12}{L^2}\bigg)-\frac{1}{4}\mathcal{F}^2-\frac{1}{4}\hat{\mathcal{F}}^2-\frac{1}{4}F_5^2-\frac{1}{4}\hat{F}_5^2\\
&
+\frac{\alpha}{3}\epsilon^{abcde}A_a \Big(3\mathcal{F}_{bc}\mathcal{F}_{de}+F^5_{bc}F^5_{de}+3 \hat{\mathcal{F}}_{bc} \hat{\mathcal{F}}_{de}+\hat{F}^5_{bc}\hat{F}^5_{de}\Big)+\frac{2\beta}{3}\epsilon^{abcde}\hat{A}_{a}\Big(3\hat{\mathcal{F}}_{bc}\mathcal{F}_{de}+\hat{F}_{bc}^{5}F_{de}^{5}\Big)\\
&
-(\hat D^{a}\Phi_{1})^{*}(\hat D_{a}\Phi_{1})-({D}^{a}\Phi_{2})^{*}({D}_{a}\Phi_{2})-V_1(\Phi_1,\Phi_2)-\frac{1}{6\eta}\epsilon^{abcde} \Big( i B_{ab}H_{cde}^*-i B_{ab}^* H_{cde}\Big)\\
&-V_2(B_{ab})-\lambda|\Phi_1|^2B_{ab}^*B^{ab}\bigg]\,,
\end{aligned}
\label{eq:action}
\eea
where $\kappa^2$ is the five dimensional gravitational constant, $L$ is the AdS radius\footnote{In the following calculations, we will set $2\kappa^2=L=1$.}. In equation \eqref{eq:1Lagrangian}, the gauge field $V_a$ represents the electromagnetic $U(1)$ current, while the axial gauge field $A_a$ represents the axial $U(1)$ current. The field strengths are $F = dV$ and $F_5 = dA$, respectively.
Here two vector gauge fields $V$, $\hat{V}$ and two axial gauge fields $A_5$, $\hat{A}_5$ are introduced based on similar consideration as in the holographic Z$_2$-Weyl semimetal \cite{Ji:2021aan}. This is different from the holographic Weyl semimetal \cite{Landsteiner:2019kxb,Landsteiner:2015lsa,Landsteiner:2015pdh} or topological nodal line semimetal \cite{Liu:2018bye,Liu:2020ymx}, where only one set of vector and axial gauge fields are introduced. Here the two sets of vector and axial gauge fields can be regarded as corresponding to two sets of U(1) and axial U(1) currents each being associated with half of the eight-component fermion. It could be regarded that 
$V$ and $A_5$ give the currents of the fermion degrees of freedom responsible for the Weyl nodes while $\hat{V}$ and $\hat{A}_5$ correspond to the $U(1)$ current and the axial $U(1)$ current of the fermion degrees of freedom responsible for the nodal line semimetal. The corresponding field strength are $\hat{F} = d\hat{V}$ and $\hat{F}_5 = d\hat{A}$, respectively. Therefore, the bulk gauge symmetry includes two sets of $U(1)_V\times U(1)_A$. $B_{ab}$ is a complex anti-symmetric two form field, which is dual to operators $\bar{\psi}\gamma^{\mu\nu}\psi$ and $\bar{\psi}\gamma^{\mu\nu}\gamma^5\psi$ on the boundary. This $B_{ab}$ is responsible for generating the nodal line for the nodal line semimetal phase and it has a mass term in $V_2(B_{ab})$ as it is not a gauge field here.

To be complete, we have included three Chern-Simons terms in the action with coupling constants $\alpha$, $\beta$ and $\frac{1}{\eta}$. The first term proportional to $\alpha$ produces the chiral anomaly term for the two sets of the $U(1)_V\times U(1)_A$ currents that we introduced above. The second Chern-Simons term proportional to  $\beta$ is the $Z_2$ axial anomaly term for fictitious spin gauge symmetry, which is not present in the current system as we will not consider the $Z_2$ currents. In the following calculations we will ignore this term. The third Chern-Simons term proportional to $\frac{1}{\eta}$ together with the mass term of the two-form field give rise to the equation of motion for the two-form field $B_{\mu\nu}$ with the self-duality relation of $B_{\mu\nu}$ taken into account \cite{Liu:2020ymx} as will explained in detail later. Two scalar fields $\Phi_1$ and $\Phi_2$ denote the mass deformations. The chiral symmetry breaks explicitly with nonzero value of the source of the two scalar fields. They are charged under the axial gauge symmetry, thus the two gauge covariant derivatives are defined as
\be
\hat D_{a}\Phi_{1}=\left(\partial_{a}-iq_1 \hat{A}_{a}\right)\Phi_{1} \,,~~~{D}_{a}\Phi_{2}=\left(\partial_{a}-iq_2{A}_{a}\right)\Phi_{2}\,,
\ee
where $q_1$ and $q_2$ are the axial charges of the scalar operators.
The scalar field potential is
 \be
 V(\Phi_1,\Phi_2)=m_1^2|\Phi_1|^2+m_1^2|\Phi_2|^2+\frac{\lambda_1}{2}\left(|\Phi_1|^4+|\Phi_2|^4\right)\,,
 \ee
{where $\lambda_1$ is the coupling of the scalar field, which characterises the number of UV degrees of freedom. Nonzero value of $\lambda_1$ indicates that the system could only be partially gapped at the IR side\cite{Landsteiner:2015lsa,Landsteiner:2015pdh,Liu:2018bye,Liu:2020ymx}.}
Two scalar bulk masses are set to be $m_1^2 L^2=m_2^2 L^2=-3$.

$H_{abc}$ is defined through the gauge field $\hat{A}_a$ and the anti-symmetric complex two form field $B_{ab}$ as
\bea
H_{abc}&=&\partial_a B_{bc}+\partial_b B_{ca}+\partial_c B_{ab}-iq_3 \hat{A}_a B_{bc}-iq_3 \hat{A}_b B_{ca}-iq_3 \hat{A}_c B_{ab}\,,
\eea
where $\eta$ is the coupling strength and $q_3$ is the axial charge of the two form field. Another potential $V_2=m_{3}^2 B_{ab}^*B^{ab}$ with $m_3$ is the mass of the two form field. As mentioned above, $B_{ab}$ is dual to operators $\bar{\psi}\gamma^{\mu\nu}\psi$ and $\bar{\psi}\gamma^{\mu\nu}\gamma^5\psi$. There is a self duality condition between these two tensor operators in field theory (i.e. $\bar{\psi}\Gamma^{\mu \nu}\Gamma^5\psi=-\frac{i}{2}\varepsilon^{\mu\nu}_{~~\alpha\beta}\bar{\psi}\Gamma^{\alpha \beta}\psi\,$). We can parameterize the field by writing the real and the imaginary part as $B_{\mu\nu}=\frac{1}{\sqrt{2}}\big(B_{+\mu\nu}+iB_{-\mu\nu}\big)$, where $B_{+\mu\nu}$ and $B_{-\mu\nu}$ are dual to $\bar{\psi}\gamma^{\mu\nu}\psi$ and $\bar{\psi}\gamma^{\mu\nu}\gamma^5\psi$, respectively.
To impose a duality condition in gravitational theory, the action of the two-form field $B_{ab}$ can be set at first order as in \eqref{eq:action}. This results in the four-dimensional components of the two-form field satisfying a complex self-duality relation directly\cite{Alvares:2011wb, Domokos:2011dn}. A similar approach was also used in building the improved holographic model of the nodal line semimetal\cite{Liu:2020ymx}. Following this recipe, we build the holographic model \eqref{eq:action} by combining the holographic Weyl semimetal and the improved holographic nodal line semimetal model. When the value of $\hat{V}, \hat{A}, \Phi_1$ and $B_{ab}$ are zero, we back to the holographic model for the Weyl semimetal\cite{Landsteiner:2015lsa,Landsteiner:2015pdh}. While with vanishing $V, A$ and $\Phi_2$ can we obtain the holographic model of the nodal line semimetal\cite{Liu:2020ymx}.

The equations of motion can be obtained by varying the metric, matter fields, and gauge fields. The complete expressions of them can be found in the Appendix \ref{app:b}.

\subsection{Zero temperature solutions for nine phases}
\label{sec:32}
%%%%%%%%%%%%%%%%%%%%%%%%%%%%
We shall focus on the zero temperature solutions which correspond to different quantum phases. The Ansatz at zero temperature is
\begin{eqnarray}\label{eq:ansatz}
\begin{aligned}
ds^2 &= -u(r)dt^2+\frac{dr^2}{u(r)}+f(r)(dx^2+dy^2)+h(r)dz^2\,,\\~~~
\Phi_1&=\phi_1(r)\,,\Phi_2=\phi_2(r)\,,~~~\\
A&=A_z(r)\,,
B_{xy}=-B_{yx}=\mathcal{B}_{xy}\,,
B_{tz}=-B_{zt}=i\mathcal{B}_{tz}\,,
\end{aligned}
\end{eqnarray} where fields $u, f, h, A_z, \mathcal{B}_{xy}, \mathcal{B}_{tz}, \phi_1$ and $\phi_2$ are functions of the radial coordinate $r$. {The non-zero components of $B_{ab}$ are selected as $B_{xy}$ and $B_{tz}$ based on the following consideration. In the gravitational theory, the complex two form field $B_{ab}$ is dynamical. The components $B_{ar}$ always vanish because of the radial gauge. Then we choose the nonzero value of $B_{ab}$ to be $B_{xy}$ due to the consideration of the symmetry of the nodal line semimetal. $B_{xy}$ also serves as the dual operator of $\bar{\Psi}\Gamma^{xy}\Psi$. The imaginary part of $B_{tz}$ is nonzero due to the duality relation $\bar{\psi}\Gamma^{\mu \nu}\Gamma^5\psi=-\frac{i}{2}\varepsilon^{\mu\nu}_{~~\alpha\beta}\bar{\psi}\Gamma^{\alpha \beta}\psi\,$. $B_{tz}$ serves as the dual operator of $\bar{\Psi}\Gamma^{\mu \nu}\Gamma^5\Psi$.} The details of the fields can be obtained by solving the eight independent equations of motion presented in Appendix \ref{app:b}. The holographic analogues of the mass terms and the source term of the tensor operators {$\bar{\Psi}\Gamma^{\mu}\Gamma^5\Psi$ and $\bar{\Psi}\Gamma^{\mu \nu}\Psi$ in combination with $\bar{\Psi}\Gamma^{\mu \nu}\Gamma^5\Psi$} are introduced in the UV boundary conditions presented here
\begin{equation}\label{eq:bcs}
 \lim_{r\rightarrow \infty}\,r\Phi_1 = M_1~,~~~\lim_{r\rightarrow \infty}\,r\Phi_2 = M_2~,~~~\lim_{r\rightarrow \infty}A_z = b\,,
 ~~~\mathop{\text{lim}}_{r\rightarrow \infty}~r^{-1}\mathcal{B}_{tz}=\mathop{\text{lim}}_{r\rightarrow \infty}~r^{-1}\mathcal{B}_{xy}=c.
\end{equation}
$M_1$ and $M_2$ correspond to the sources of the dual scalar operator $\Psi^{\dagger}\Gamma^{0}\mathbb{M}_1\Psi$ and $\Psi^{\dagger}\Gamma^{0}\mathbb{M}_2\Psi$, {thus they represent the mass terms of the fermions.} $b$ corresponds to the source of the chiral current $\Psi^{\dagger}\Gamma^{0}\Gamma^{\mu}\Gamma^{5}\Psi$ and $c$ corresponds to the source of $\Psi^{\dagger}\Gamma^{0}\Gamma^{\mu\nu}\Psi$ as well as $\Psi^{\dagger}\Gamma^{0}\Gamma^{\mu\nu}\Gamma^{5}\Psi$, respectively. Compared with the weakly coupled Lagrangian, the boundary values $c$ and $M_1$ here should be responsible for the nodal line semimetal, i.e. a certain combination of them determines the effective radius of the nodal ring. Similarly, the boundary values of $b$ and $M_2$ are responsible for the Weyl nodes, and a combination of them shows the effective distance between the two Weyl nodes.
To analyze the coexistence of a Weyl semimetal and a nodal line semimetal in the strong coupling limit, we can solve this holographic system by activating these four sources. There are nine different types of IR solutions that can be found, which flow to the asymptotic AdS$_{5}$ boundary to give nine types of full spacetime solutions. From the IR behavior of these nine solutions along with additional supporting evidence from anomalous conductivities, we can identify nine distinct phases corresponding to these solutions.  From the boundary value of each IR solution, we can determine a phase diagram similar to the weakly coupled one in Fig.\ref{fig:01}. There are four phases ({i.e., the Weyl-nodal
phase, the gap-nodal phase, the Weyl gap phase and the gap-gap phase) with four phase boundaries (i.e., the Weyl-Dirac line, the gap-Dirac line, the Dirac-gap
line and the Dirac-nodal line.) and one double critical point in the $\hat{M}_1-\hat{M}_2$ plane of the phase diagram. These nine solutions and their corresponding phases are listed in order below.
\vspace{.4cm}\\
\emph{\textbf{The Double Critical Point}}~~
The double critical solution corresponds to the state where both of the nodal ring and the Weyl nodes form a critical Dirac node in the weakly coupled theory (Fig.~\ref{fig:eight}(d)). The corresponding geometry in the holographic system is a Lifshitz-type solution. {As in the case in the Weyl semimetal and the nodal line semimetal, the nontrivial Lifshitz-type solution has no free parameter at the IR side, thus it is unique and leads to the only one critical value at the boundary, which corresponds to the double critical phase transition point. } The solutions is
\begin{eqnarray}
\begin{aligned}
\label{eq:nhcri}
&ds^{2}=u_{0}r^{2}\left(-dt^{2}+dx^{2}\right)+\frac{dr^{2}}{u_{0}r^{2}}+f_{0}r^{\alpha}dy^{2}+h_{0}r^{2\alpha_{1}}dz^{2}\,,\\
&A_{z}=r^{\alpha_{1}}\,,~~\phi_{1}=\phi_{10}\,,~~\phi_{2}=\phi_{20}\,,\\
&\mathcal{B}_{xy}=b_{xy}^{\left(c\right)} r^{\alpha}\,,~~\mathcal{B}_{tz}=b_{tz}^{\left(c\right)}r^{1+\alpha_1}\,,
\end{aligned}
\end{eqnarray}
where $\left\{ u_{0},f_{0},h_{0},\alpha,\alpha_{1},\phi_{10},\phi_{20},b_{tzc}\right\}$ are constants.
This solution has an anisotropic Lifshitz-type symmetry $\left(t,x\right)\rightarrow s\left(t,x\right),~y\to s^{\alpha/2}y,~z\to s^{\alpha_{2}}z$. The constants $\left\{ u_{0},f_{0},h_{0},\alpha,\alpha_{1},\phi_{10},\phi_{20},b_{tz}^{\left(c\right)}\right\}$ in (\ref{eq:nhcri}) can be fully determined by the values of the parameters $m_1, m_2, m_3, q_1, q_2, q_3, \lambda, \lambda_1, \eta$. The value of $b_{xy}^{\left(c\right)}$ can be set to 1 based on the scaling symmetry of the $x-y$ plane. The fact that this solution is the double critical point can be further supported from the IR behavior of the fields $A_z^5$ and $B_{xy}$, where they are IR irrelevant and the IR behavior of these operators will completely determine the corresponding phase of the solution as summarized in section 4 of \cite{Liu:2018bye}. Similarly the corresponding phases of the other eight solutions in this system can also be found from the IR behaivor of the corresponding fields, i.e. $A_z^5$, $B_{xy}$, $B_{tz}$, $\Phi_1$ and $\Phi_2$ etc. using the general paradigm proposed in \cite{Liu:2018bye}.

As \eqref{eq:nhcri} is an exact solution, irrelevant deformations of the geometry should
be introduced to ensure that it flows to asymptotic AdS in the UV. In the IR the leading order solutions with irrelevant perturbations are
\begin{eqnarray}
\label{eq:nh-criticalsol}
\begin{aligned}
u&=u_{0}r^{2}\left(1+\delta u_{1}\,r^{\beta_{1}}+\delta u_{2}\,r^{\beta_{2}}\right)\,,\\
f&=f_{0}r^{\alpha}\left(1+\delta f_{1}\,r^{\beta_{1}}+\delta f_{2}\,r^{\beta_{2}}\right)\,,\\
h&=h_{0}r^{2\alpha_{1}}\left(1+\delta h_{1}\,r^{\beta_{1}}+\delta h_{2}\,r^{\beta_{2}}\right)\,,\\
A_{z}&=r^{\alpha_{1}}\left(1+\delta a_{1}\,r^{\beta_{1}}+\delta a_{2}\,r^{\beta_{2}}\right)\,,\\
\mathcal{B}_{tz}&=b_{tz}^{\left(c\right)}r^{1+\alpha_1}\left(1+\delta \mathcal{B}_{tz1}\,r^{\beta_{1}}+\delta \mathcal{B}_{tz2}\,r^{\beta_{2}}\right)\,,\\
\mathcal{B}_{xy}&=r^{\alpha}\left(1+\delta \mathcal{B}_{xy1}\,r^{\beta_{1}}+\delta \mathcal{B}_{xy2}\,r^{\beta_{2}}\right)\,,\\
\Phi_{1}&=\phi_{10}\left(1+\delta\phi_{11}\,r^{\beta_{1}}+\delta\phi_{12}\,r^{\beta_{2}}\right)\,,\\
\Phi_{2}&=\phi_{20}\left(1+\delta\phi_{21}\,r^{\beta_{1}}+\delta\phi_{22}\,r^{\beta_{2}}\right)\,.
\end{aligned}
\end{eqnarray}

Taking into account the scaling symmetries of the system, there are only two free parameters above: $\delta a_{1}$ and $\delta \phi_{11}$. The boundary values of $c/b$ and $M_{1}/c$ then are determined by the IR values of $\delta a_{1}$ and $\delta \phi_{11}$. This is different from the case in the holographic Weyl semimetal system, which has no free parameter for the critical solution as the system has a fixed critical point. It is also different from the case in the $Z_2$-Weyl semimetal where only one free parameters controls the value of $c/b$. In this system, the existence of two free parameters transforms the single critical point into a critical plane with varying values of $c/b$ and $M_{2}/c$ in the three dimensional phase diagram. 
Varying the values of $c/b$ and $M_{2}/c$ will determine the corresponding critical values of $M_{1c}/b$.

Without loss of generality, we fix $\lambda=1$, $\lambda_1=1/10$, $q_1=1/2$, $q_2=3/2$, $q_3=1/4$, $m_3=1$ and $\eta=2$ in the numerics. We then could find that 
$\{u_0, f_0, h_0, \phi_{10}, \phi_{20}, b_{tzc}, \alpha, \alpha_1, \beta_1, \beta_2\}\to \{4.286,\,0.732,\,0.770,\,0.727,\,0.879, \,1.142,
\,0.339, \,0.606, \,0.736, \,1.189\}$,
$\{\delta u_{1}, \delta f_{1}, \delta h_{1}, \delta \mathcal{B}_{tz1},\\ \delta \mathcal{B}_{xy1}, \delta\phi_{11}, \delta\phi_{21}\}\to$ \{4.443, \,-3.104,\, -14.48,\, -2.183, \,-0.369,\, 1.909, \,5.795\}$\delta a_1$ and $\{\delta u_{2}, \delta f_{2}, \delta h_{2}, \\ \delta a_{2}, \delta \mathcal{B}_{tz2}, \delta \mathcal{B}_{xy2}, \delta\phi_{22}\}\to\{-1.007,\,2.658,
\,-1.802,\,-0.153,\,-2.235,\,0.365,\,0.277\}\delta\phi_{11}$. Shooting to $c/b=1$ we have the critical values $\hat{M}_1\equiv \frac{M_1}{c}=1.073$ and $\hat{M}_2\equiv \frac{M_2}{b}=0.924$, which corresponds to the double critical Dirac point in the phase diagram, and has a similar configuration as the weak coupling field theory of \eqref{eq:1Lagrangian}. This double critical solution is marked by the red point in the phase diagram Fig.\ref{fig:02}.

In the following, we will introduce four additional solutions for the four phase boundaries. In these phases, the Weyl nodes (or the nodal ring) become a Dirac node, while the nodal ring (or the Weyl nodes) still survive or becomes as gap state. These phases are represented by dashed lines in the phase diagram shown in Fig.\ref{fig:02}.
\vspace{.4cm}\\
\emph{\textbf{Weyl-Critical}}~~
The Weyl-critical phase refers to the phase boundary where the nodal line semimetal becomes critical while the two Weyl nodes still survive, similar to the weak coupling field theory case (Fig.~\ref{fig:eight}(b)). The holographic solution in the IR takes the following form
\begin{eqnarray}
\begin{aligned}
&u=u_{0}\,r^{2}\left(1+ \delta u_{1}\,r^{\alpha_{1}}\right)\,,\\
&f=f_{0}\,r^{2\alpha}\left(1+ \delta f_{1}\,r^{\alpha_{1}}\right)\,,\\
&h=h_{0}\,r^{2}\left(1+ \delta h_{1}\,r^{\alpha_{1}}\right)\,,\\
&A_{z}=a_{0}+\phi_{20}^{2}r^{1-\alpha}h_{0}\exp\left(-\frac{3a_{0}}{r\sqrt{u_{0}h_{0}}}\right)\,,\\
&\mathcal{B}_{tz}=b_{tz0}r^{2}\left(1+\delta \mathcal{B}_{tz1}\,r^{\alpha_{1}}\right)\,,\\
&\mathcal{B}_{xy}=r^{\alpha}\left(1+\delta \mathcal{B}_{xy1}\,r^{\alpha_{1}}\right)\,,\\
&\Phi_{1}=\phi_{10}\left(1+\delta\phi_{11}\,r^{\alpha_{1}}\right)\,,\\
&\Phi_{2}=\phi_{20}\exp\left(-\frac{3a_{0}}{2r\sqrt{u_{0}h_{0}}}\right)r^\frac{-1-\alpha}{2}\,\,.
\label{wc}
\end{aligned}
\end{eqnarray}

With the above choice of $q_1,q_2, q_3, \lambda, \lambda_1,\eta$, we have $\{u_0, f_0, \phi_{10},\alpha, \alpha_1\}$
$\to \{2.735$,\, $0.754$,
$\,0.557$,\, $0.314$,\, $1.274\}$ and $\{\delta u_{1}, \delta f_{1}, \delta h_{1}, \delta \mathcal{B}_{tz1}, \delta \mathcal{B}_{xy1},\delta\phi_{11}\} \to \{1.399,\, -3.411,\,1.399,\, 2.723,\\\,-0.402, \,1.585\}$. Three free parameters are $h_0$, $a_0$ and $\phi_{20}$. $\mathcal{B}_{tz0}=0.869\sqrt{h_0}$. With these three shooting parameters, we can fix $c/b$ and get a curve in the plane $M_1/c$-$M_2/b$, i.e. a phase boundary in the plane phase diagram. Shooting to $c/b=1$ we obtain the (dashed cyan in Fig.\ref{fig:02}) curve as the critical line with two endpoints being (1.073,~0.924) and (0.858,~0) in the phase diagram of the $\hat{M}_1$-$\hat{M}_2$ plane. The first end point is exactly the double critical point obtained earlier. The other end point (0.858,~0) is consist with the critical value in the holographic model of single nodal line semimetal\cite{Liu:2020ymx}.
\vspace{.4cm}\\
\emph{\textbf{Critical-Nodal}}~~
This kind of solution corresponds to the phase boundary where two Weyl nodes annihilate into a Dirac node while the nodal ring still exists, similar to the weak coupling field theory case of Fig.~\ref{fig:eight}(e). As mentioned in \cite{Liu:2018bye,Liu:2020ymx}, the geometry of the nodal line is an emergent Lifshitz-type symmetry in the deep IR region. Combine with the critical Lifshitz-type solution of the Weyl semimetal case, we can obtain the corresponding solution with the following form(with the same choice of $q_1,q_2, q_3, \lambda, \lambda_1,\eta$):
\begin{eqnarray}
\begin{aligned}
&u=u_{0}\,r^{2}\left(1+ \delta u_{1}\,r^{\alpha_{1}}\right)\,,\\
&f=f_{0}\,r^{2\alpha}\left(1+ \delta f_{1}\,r^{\alpha_{1}}\right)\,,\\
&h=h_{0}\,r^{2\alpha_{2}}\left(1+ \delta h_{1}\,r^{\alpha_{1}}\right)\,,\\
&A_{z}=r^{\alpha_2}\left(1+ \delta a_{1}\,r^{\alpha_{1}}\right)\,,\\
&\mathcal{B}_{tz}=b_{tz0}r^{1+\alpha_2}\left(1+\delta \mathcal{B}_{tz1}\,r^{\alpha_{1}}\right)\,,\\
&\mathcal{B}_{xy}=r^{\alpha}\left(1+\delta \mathcal{B}_{xy1}\,r^{\alpha_{1}}\right)\,,\\
&\Phi_{1}=\phi_{10}r^{\beta}\,,\\
&\Phi_{2}=\phi_{20}\left(1+ \delta \phi_{21}\,r^{\alpha_{1}}\right)\,\,.
\label{nc}
\end{aligned}
\end{eqnarray}
Here for the same parameter values of the system, numerically we can obtain $\{u_0, f_0, h_0, b_{tz0}, \alpha, \alpha_1,\,\beta\}$
$\to \{2.727$,\,$0.635$,
$\,2.583$,\,$1.265$,\,$0.183$,\,$1.234$,\,$0.228\}$ and $\{\delta u_{1}, \delta f_{1},\\ \delta h_{1}, \delta \mathcal{B}_{tz1}, \delta \mathcal{B}_{xy1},\delta\phi_{21}\} \to \{0.712,\, -5.250,\,0.763,\, 1.679,\,-0.054, \,2.835\}\delta a_{1}$. The other two free parameters are $\phi_{10}$ and $\phi_{20}$. With these shooting parameters, the value of $c/b$ can be fixed to 1 and we obtain the (dashed blue line in Fig.\ref{fig:02}) curve as the critical line with two endpoints being (1.073,~0.924) and (0.731,~0) in the phase diagram of the $\hat{M}_1$-$\hat{M}_2$ plane. The first endpoint is again the double critical pint while the second endpoint (0.731,~0) is consistent with the critical value in the holographic model of single Weyl semimetal\cite{Landsteiner:2015pdh}.
\vspace{.4cm}\\
\emph{\textbf{Gap-Critical}}~~
This is the solution that corresponds to the phase boundary where two Weyl nodes become a gapped trivial phase while the radius of the nodal ring becomes zero, similar to the weak coupling field theory in Fig.~\ref{fig:eight}(h). The bulk solution in the IR is
\begin{eqnarray}
\begin{aligned}
&u=u_{0}\,r^{2}\left(1+\delta u_{1}\,r^{\alpha_{1}}+ \,\delta u_2\, r^{2 \alpha_3-2}\right)\,,\\
&f=f_{0}\,r^{\alpha}\left(1+\delta f_{1}\,r^{\alpha_{1}}
+\, \delta f_2\, r^{2 \alpha_3 -2}\right)\,,\\
&h=u_{0}\,r^{2}\left(1+\delta h_{1}\,r^{\alpha_{1}}+ \,\delta h_2\, r^{2 \alpha_3-2}\right)\,,\\
&A_{z}=a_{0}\,r^{\alpha_3}\,, \\
&\mathcal{B}_{tz}=\mathcal{B}_{tz0}r^2\left(1+\delta \mathcal{B}_{tz1}\,r^{\alpha_{1}}
+\, \delta \mathcal{B}_{tz2}\, r^{2 \alpha_3 -2}\right)\,,\\
&\mathcal{B}_{xy}=\mathcal{B}_{xy0}r^{\alpha}\left(1+\delta \mathcal{B}_{xy1}\,r^{\alpha_{1}}
+\, \delta \mathcal{B}_{xy2}\, r^{2 \alpha_3 -2}\right)\,,\\
&\Phi_{1}=\phi_{10}\left(1+\delta \phi_{11}\,r^{\alpha_{1}}
+\, \delta \phi_{12}\, r^{2 \alpha_3 -2}\right)\,,\\
&\Phi_{2}=\sqrt{30}\left(1+\delta \phi_{21}r^{\alpha_2}+r^{\frac{1}{2} \left(-\alpha +\frac{\sqrt{(\alpha +2)^2 \text{u0}+24}}{\sqrt{\text{u0}}}-2\right)}\right)\,\,.
\label{ct}
\end{aligned}
\end{eqnarray}
In \eqref{ct}, for the system parameters that we have chosen, we have $\alpha=0.647$, $u_0=11.398$, $f_0=0.705\mathcal{B}_{xy0}$, $\mathcal{B}_{tz0}=9.186$, $\alpha_1=0.988$, $\alpha_2=0.186$ and $\alpha_3=3.133$,. The other parameters are $\{\delta u_{1},\delta f_{1}, \delta h_{1}, \delta \mathcal{B}_{tz1}, \delta \mathcal{B}_{xy1}, \delta \phi_{11}\} \to \{-0.074,\,0.208,\, 2.297,\,-0.157,\,0.048,\,-0.026\}\delta \phi_{21}$ and $\{\delta u_{2},\delta f_{2}, \delta h_{2}, \delta \mathcal{B}_{tz2}, \delta \mathcal{B}_{xy2}, \delta \phi_{12}\} \to \{0.009,\,0.011,\, -0.054,\,-0.012,\,0.002,\,-0.0003\}a_0$. This solution exists only when $M_1>1.073$ and $M_2>0.924$, as shown by the bottom purple dotted lines in the phase diagram in Fig. \ref{fig:02}. 
\vspace{.4cm}\\
\emph{\textbf{Critical-Gap}}~~
This is the phase that corresponds to the annihilation of two Weyl nodes into one Dirac node, while the ring of nodes becomes a gap, similar to the weakly coupled case in Fig.~\ref{fig:eight}(f). The corresponding solution in the holographic model has the following form
\begin{eqnarray}
\begin{aligned}
&u=u_{0}\,r^{2}\left(1+\delta u_{1}\,r^{\alpha_{1}}\right)\,,\\
&f=r^{2}\left(1+\delta f_{1}\,r^{\alpha_{1}}\right)\,,\\
&h=h_{0}\,r^{2\alpha}\left(1+\delta h_{1}\,r^{\alpha_{1}}\right)\,,\\
&A_{z}=r^{\alpha}\left(1+\delta a_{1}\,r^{\alpha_{1}}\right)\,,\\
&\mathcal{B}_{tz}=\mathcal{B}_{tz0} r^{\alpha_{2}}\,,\\
&\mathcal{B}_{xy}=\mathcal{B}_{xy0} r^{\alpha_{2}}\,,\\
&\Phi_{1}=\sqrt{30}\left(1+\delta \phi_{11}\,r^{\beta}\right)\,,\\
&\Phi_{2}=\phi_{20}\left(1+\delta \phi_{21}r^{\alpha_1}\right)\,.
\label{tc}
\end{aligned}
\end{eqnarray}
The shooting parameters in (\ref{tc}) are $\mathcal{B}_{xy0}$, $\delta a_{1}$ and $\phi_{20}$. Other parameters can be determined by the equations of motion as $\{u_0, h_0, \mathcal{B}_{tz0}, \phi_{20},\alpha,\alpha_1\,\alpha_2,\beta,\} \to \{6.040,\,0.818,\,6.040\mathcal{B}_{xy0},\,\\1.581,\,0.621,\,0.407,\,14.223,\,0.256\}$ and $\{\delta u_{1}, \delta f_{1}, \delta h_{1}, \delta \phi_{11}\} \to \{0.021,\,-0.158,\, 0.343,\,0.057\}\delta a_{1}$. When $M_1>1.073$ and $M_2>0.924$, shooting to $c/b=1$ results in another purple dotted line, as shown in the phase diagram in Fig. \ref{fig:02}.

Finally, we will present the rest four solutions that correspond to the rest four regions in the phase diagram Fig.\ref{fig:02}. 
\vspace{.4cm}\\
\emph{\textbf{Weyl-Nodal}}~~
This solution  corresponds to the phase where the Weyl semimetal and the nodal line semimetal coexist, similar to the weakly coupled case in Fig.~\ref{fig:eight}(a). In holography, the corresponding geometry in the IR can be obtained by comparing with the forms of the Weyl semimetal state and the nodal line semimetal state in the holographic model, which is:
\begin{eqnarray}
\begin{aligned}
&u=u_{0}\,r^{2}\left(1+\delta u\,r^{\alpha_{1}}\right)\,,\\
&f=f_{0}\,r^{\alpha}\left(1+\delta f\,r^{\alpha_{1}}\right)\,,\\
&h=r^{2}\left(1+\delta h\,r^{\alpha_{1}}\right)\,,\\
&A_{z}=a_0+\exp\left(-\frac{3 a_0}{r \sqrt{u_0}}\right)r^{\alpha-1}\,,\\
&\mathcal{B}_{tz}=\mathcal{B}_{tz0} r^{2}\left(1+\delta \mathcal{B}_{tz}\,r^{\alpha_{1}}\right)\,,\\
&\mathcal{B}_{xy}=r^{\alpha}\left(1+\delta \mathcal{B}_{xy}\,r^{\alpha_{1}}\right)\,,\\
&\Phi_{1}=\phi_{10}\,r^{\beta}\,,\\
&\Phi_{2}=\phi_{20}\exp\left(-\frac{3 a_0}{2 r \sqrt{u_0}}\right)r^{-\frac{1+\alpha}{2}}\,.
\label{wn}
\end{aligned}
\end{eqnarray}
There are four shooting parameters $\delta u$, $a_0$, $\phi_{10}$ and $\phi_{20}$ in (\ref{wn}). With the choice of $\lambda=1$, $\lambda_1=1/10$, $q_1=1/2$, $q_2=3/2$, $q_3=1/4$, $m_3=1$ and $\eta=2$, we have $\{u_0, f_0, \mathcal{B}_{tz0}, \alpha, \alpha_1, \beta\}\to$\{2.727, \, 0.635, \, 0.787,\,0.183,\, 1.273,\, 0.228\},
$\{\delta f, \delta h, \delta \mathcal{B}_{tz},\delta \mathcal{B}_{xy}\}$
$\to$\{-2.616, 1, 1.719, -0.302\}$\delta u$. When shooting to $c/b=1$, this solution only exists at the lower-left part of the phase diagram of Fig.~\ref{fig:02} between two phase boundaries.
\vspace{.4cm}\\
\emph{\textbf{Weyl-Gap}}~~
This kind of solution corresponds to the phase where the nodal line semimetal becomes gapped and the Weyl semimetal still exists as in Fig.~\ref{fig:eight}(c). Comparing with the topological trivial geometry in the holographic nodal line semimetal, we can obtain the corresponding holographic solution for the Weyl-gap phase is
\begin{eqnarray}
\begin{aligned}
&u=u_0\,r^{2}\,,\\
&f=h=r^{2}\,,\\
&A_{z}=a_0+\exp\left(-\frac{3 a_0}{r \sqrt{u_0}}\right)\left(\frac{\phi_{20}^2 u_0}{9 \pi  a_{0}^3}+\frac{\phi_{20}^2 \sqrt{u_0}}{3 \pi  a_{0}^2 r}\right)\,,\\
&\mathcal{B}_{tz}=\mathcal{B}_{tz0} r^{\beta}\,,\\
&\mathcal{B}_{xy}=\mathcal{B}_{xy0}r^{\beta}\,,\\
&\Phi_{1}=\phi_{10}+\phi_{11}\,r^{\alpha}\,,\\
&\Phi_{2}=2\phi_{20}\frac{\exp\left(-\frac{3 a_0}{2 r \sqrt{u_0}}\right)}{\sqrt{3 \pi } r^2 \sqrt{\frac{a_0}{r \sqrt{u_0}}}}\,.
\label{wg}
\end{aligned}
\end{eqnarray}
The parameters in (\ref{wg}) are $\{u_0, \mathcal{B}_{tz0},\phi_{10},\alpha, \beta\}\to\{4.75, \, 2.179\mathcal{B}_{xy0}, \,5.47723,\,0.294,\, 14.224\}$. Four shooting parameters are $\mathcal{B}_{xy0}$, $a_0$, $\phi_{11}$ and $\phi_{20}$. The region of this solution is the lower-right part of the phase diagram in Fig.~\ref{fig:02}.
\vspace{.4cm}\\
\emph{\textbf{Gap-Nodal}}~~
This solution corresponds to the phase where the two Weyl nodes annihilate to form a gap while the nodal line semimetal still exists, as in Fig.~\ref{fig:eight}(g). Similar to the above, we can obtain the holographic solution for the gap-nodal phase by combining the topological trivial geometry in the holographic Weyl semimetal state with the Lifshitz-type solution for the nodal line semimetal
\begin{eqnarray}
\begin{aligned}
&u=u_{0}\,r^{2}\left(1+ \delta u_{1}\,r^{\alpha_{1}}\right)\,,\\
&f=f_{0}\,r^{\alpha}\left(1+ \delta f_{1}\,r^{\alpha_{1}}\right)\,,\\
&h=r^{2}\left(1+ \delta h_{1}\,r^{\alpha_{1}}\right)\,,\\
&A_{z}=a_0 r^{\alpha_2}\,,\\
&\mathcal{B}_{tz}=b_{tz0}r^{2}\left(1+\delta \mathcal{B}_{tz1}\,r^{\alpha_{1}}\right)\,,\\
&\mathcal{B}_{xy}=r^{\alpha}\left(1+\delta \mathcal{B}_{xy1}\,r^{\alpha_{1}}\right)\,,\\
&\Phi_{1}=\phi_{10}r^{\beta}\,,\\
&\Phi_{2}=\phi_{20}+\phi_{21}\,r^{\alpha_{3}}\,.
\label{gn}
\end{aligned}
\end{eqnarray}
The corresponding parameters are $\{u_0, f_0, \mathcal{B}_{tz0}, \phi_{20}, \alpha, \alpha_1,\alpha_2,\alpha_3, \beta\}\to$\{13.995,\,-0.270,\,-1.854,\,0.036,\, 1.468,\,-2.228,\, 3.088,\, 0.641\},
$\{\delta f_1, \delta h_1, \delta \mathcal{B}_{tz},\delta \mathcal{B}_{xy}\}$
$\to$\{-2.466, 1, 1.679, -0.054\}$\delta u_1$. Four shooting parameters are $a_0$, $\delta u_1$, $\phi_{10}$ and $\phi_{21}$, which can be tuned to fix the radio  $c/b=1$. This solution only exists at the upper-left part of the phase diagram of Fig.~\ref{fig:02}.
\vspace{.4cm}\\
\emph{\textbf{Gap-Gap}}~~
This kind of solution corresponds to the topological trivial state with no semimetal states, as in Fig.~\ref{fig:eight}(h). The IR geometry for the gap-gap phase is
\begin{eqnarray}
\begin{aligned}
&u=u_{0}\,r^{2}\,,\\
&f=h=r^{2}\,,\\
&A_z=a_0 r^{\alpha_1},\\
&\mathcal{B}_{tz}=\mathcal{B}_{tz0} r^{\beta}\,,\\
&\mathcal{B}_{xy}=\mathcal{B}_{xy0}r^{\beta}\,,\\
&\Phi_{1}=\phi_{10}+\phi_{11}\,r^{\alpha_2}\,,\\
&\Phi_{2}=\phi_{20}+\phi_{21}\,r^{\alpha_2}\,.
\label{gg}
\end{aligned}
\end{eqnarray}
In (\ref{gg}), we have $u_0=8.5$, $\mathcal{B}_{xy0}=0.343\mathcal{B}_{tz0}$,$\phi_{10}=\phi_{20}=5.477$, $\alpha_1=3.109$, $\alpha_2=0.169$ and $\beta=10.633$ with the choice of $\lambda=1$, $\lambda_1=1/10$, $q_1=1/2$, $q_2=3/2$, $q_3=1/4$, $m_3=1$ and $\eta=2$. The shooting parameters are $\mathcal{B}_{tz0}$, $a_0$, $\phi_{10}$ and $\phi_{20}$. This solution only exists at the upper-right part of the phase diagram of Fig.~\ref{fig:02}.

\vspace{0cm}
\begin{figure}[ht!]
  \centering
\includegraphics[width=0.8\textwidth]{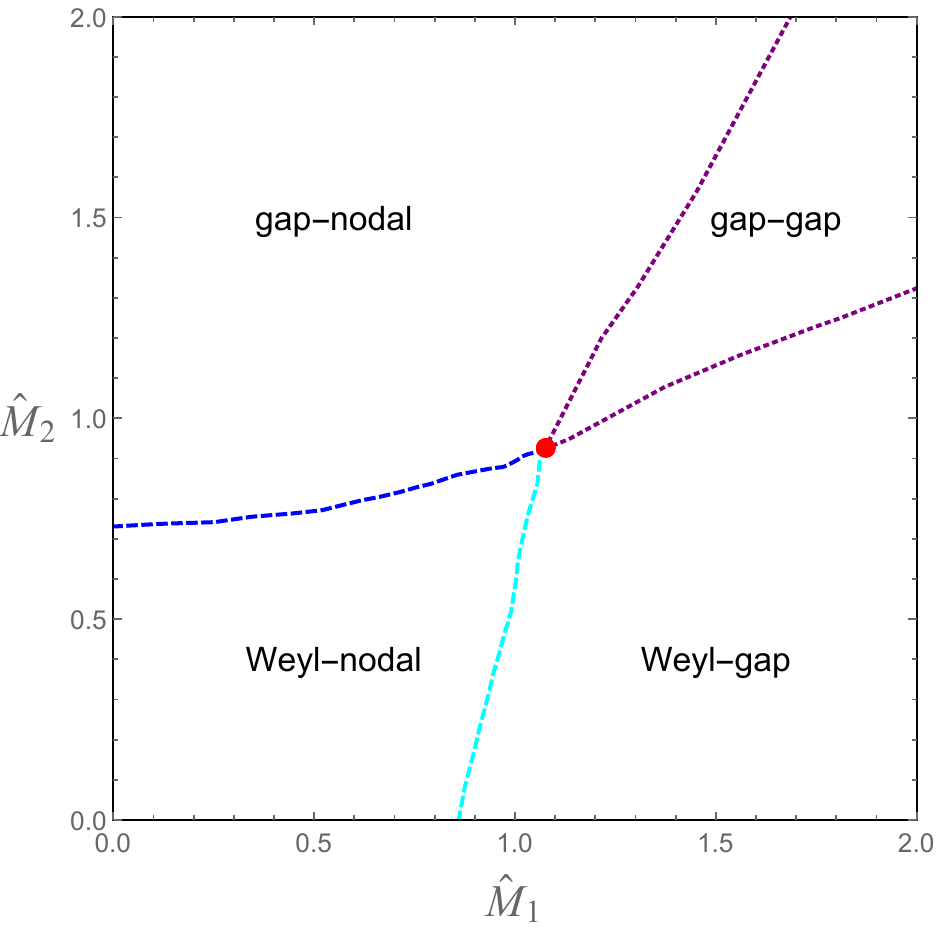}
\vspace{-0.3cm}
  \caption{\small The phase diagram of the holographic system \eqref{eq:action} with three dimensionless parameters $\hat{M}_1=M_1/c$, $\hat{M}_2=M_2/b$ and $c/b=1$. The red point is the double critical point at which both Weyl nodes and the nodal ring become critical. The blue dashed line corresponds to the critical phase in which the Weyl nodes annihilate into a critical Dirac node while a nodal ring still exists. The cyan dashed line corresponds to the critical phase in which the radius of the nodal ring becomes zero while a pair of Weyl nodes still exists. The purple dotted lines correspond to the phase where the pair of Weyl nodes annihilates into a critical Dirac point (or the radius of the nodal ring becomes zero) while the nodal line (or the pair of Weyl nodes) becomes gapped.}
 \label{fig:02}
\end{figure}

By examining the entire region where the solutions above exist, we can construct a two-dimensional phase diagram for the holographic model with a fixed ratio of c/b = 1, as shown in Fig.\ref{fig:02}.
The red point represents the double critical point, while the four curved lines correspond to the phase boundaries where one single critical Dirac node is formed. These critical lines separate four regions that correspond to four different phases, similar to the weakly coupled field theory. In comparison to the phase diagram in the weak coupled case in Fig.\ref{fig:01}, the range of each phase is quite different in the strongly coupled system, especially in the region near the critical point. The system can undergo phase transitions between the gap-nodal phase and the Weyl gap phase by fixing the value of either $M_1$ or $M_2$ while changing the other near the critical point. This is a phenomenon that is not present in the weak coupling model. The phase transitions should be continuous ones similar to the cases of holographic 
Weyl and nodal line semimetals. We will see more details on this point in the next subsection from the free energy. It should be noted that the value of $c/b$ could be arbitrary, resulting in a three-dimensional phase diagram, similar to the case of the $Z_2$-Weyl semimetal\cite{Ji:2021aan}.

Note that we have identified each phase with the solutions from the IR behavior of these IR solutions, as the IR behavior determines the low energy behavior of the corresponding state as pointed out in \cite{Liu:2018bye,Liu:2020ymx}. The nontrivial IR relevant values of the $b_z$ and $b_{xy}$ fields in this case imply nontrivial expectation values of the corresponding operators at low energy. Thus bulk solutions with nontrivial IR behavior in $b_z$ (and/or $b_{xy}$) would correspond to nontrivial Weyl nodes (and/or nodal ring) at the boundary \cite{Liu:2018bye,Liu:2020ymx}. This point was confirmed in both \cite{Landsteiner:2015lsa,Landsteiner:2015pdh} and \cite{Liu:2018bye,Liu:2020ymx} by the calculation of the anomalous Hall conductivity, which is the order parameter for a Weyl semimetal topological phase transition, and by calculating the fermionic spectral functions for a topological nodal line semimetal to reveal the nodal rings directly. Topological invariants were also obtained as further evidence for both systems in \cite{Liu:2018djq}. Thus here in this work, we can directly read the corresponding phase from the IR solutions. Nevertheless we will also calculate the corresponding anomalous Hall conductivity and free energy for this system in the next subsection, which will provide further consistency check and provide more details on the continuous behavior of the phase transitions. Further calculations on the fermionic spectral functions and topological invariants will be left to a future work.
\subsection{Anomolous Hall Conductivity and Free Energy}
\label{sec:33}
In this subsection, we will focus on the phase transition process between the gap-nodal and Weyl-gap phases in Fig.\ref{fig:02}. This novel phase transition process cannot be present in a model where only a single topological semimetal state survives. As shown in Fig.\ref{fig:02}, by fixing either $M_1$ or $M_2$ near the critical point and changing the other dimensionless mass parameter can cause the system to undergo a phase transition between the gap-nodal and Weyl-gap phases. In contrast to the weak coupling case shown in Fig.\ref{fig:01}, this phase transition process can only be realized by simultaneously changing both $M_1$ and $M_2$.

There is no simple order parameter for the nodal line topological phase transition, therefore, in this subsection we calculate the anomalous Hall conductivity as part of the order parameter. In the following, the anomalous Hall conductivity will be calculated with the value of $M_1$ fixed to focus on the phase transition between the Weyl semimetal phase and the trivial gapped phase. The anomolous hall conductivity can be computed by the Kubo formula
\begin{equation}
\label{ahe}
 \sigma_{xy} = \lim_{\omega\rightarrow 0}\frac{1}{i\omega} \langle J_x J_y \rangle (\omega,\vec k =0)\,.
\end{equation}
Eq.\eqref{ahe} is the retarded correlation function and can be calculated through the fluctuations of the gauge fields
around the background in holography, where the in-falling boundary conditions need to be taken into consideration. The final result of the anomalous Hall conductivity is determined by the horizon value of the $U(1)$ gauge field, i.e. $\sigma_{\text{AHE}}=8\alpha A_z\left(r_0\right)$, where $r_0$ is the location of the horizon. Details of the calculation can be found  in\cite{Landsteiner:2015pdh,Ji:2021aan} and here in our system, extra fields will not affect this result as the fluctuations of the fields are decoupled.

\vspace{0cm}
\begin{figure}[ht!]
  \centering
\includegraphics[width=0.6\textwidth]{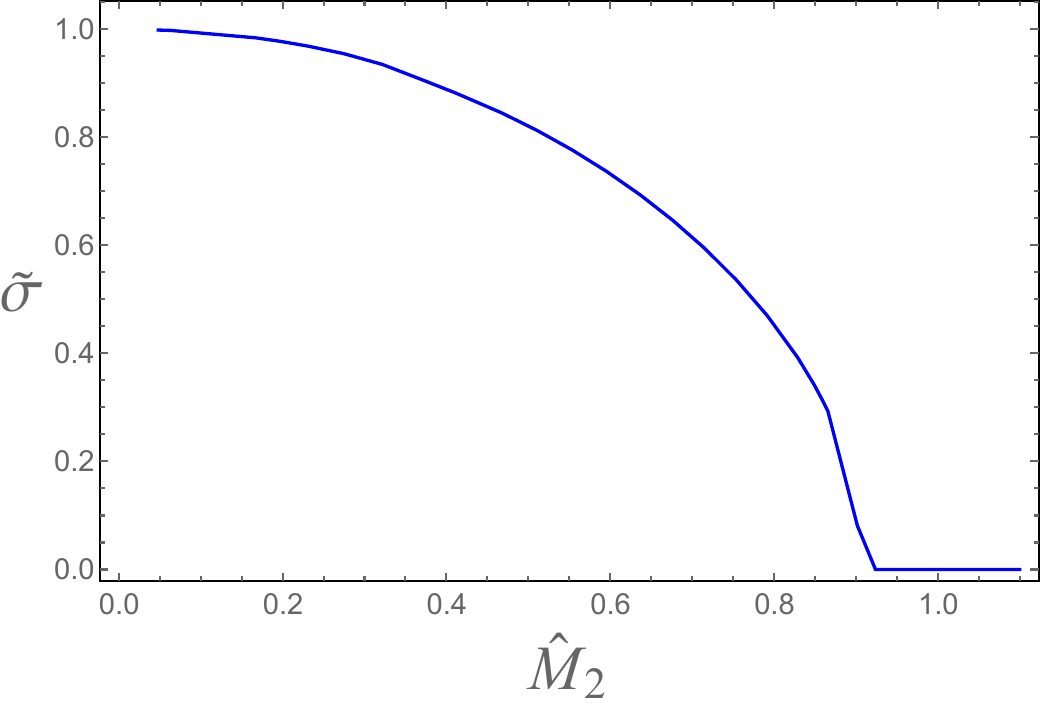}
\vspace{-0.3cm}
  \caption{\small The anomalous Hall conductivity as a function of $\hat{M}_2$ at $c/b =1$ and $\hat{M}_1=1.073$ at zero temperature. The critical behavior of the anomalous Hall conductivity near the phase transition point is $\tilde{\sigma}\propto(\hat{M}_{2c}-\hat{M}_{2})^{\alpha}$, with $\alpha \approx 0.409$.} 
 \label{fig:03}
\end{figure}
In Fig.\ref{fig:03} we plot the anomalous Hall conductivity as a function of $\hat{M}_2$ at zero temperature with fixed value $\hat{M}_1=1.073$ and $c/b =1$, i.e. the curve will pass through the double critical point, which is at $\hat{M}_1$=1.073 and $\hat{M}_2$=0.924. When $\hat{M}_2<0.924$, the system is in the Weyl-gap phase while when $\hat{M}_2>0.924$, the system will be in the gap-nodal phase. As shown in Fig.\ref{fig:03}, the value of the anomalous Hall conductivity is non-zero when $\hat{M}_2<0.924$, i.e. in the Weyl-gap phase.  When $\hat{M}_2$ larger than $0.924$, it becomes zero in the gap-nodal phase\footnote{The anomalous Hall conductivity exhibits a smooth behavior but experiences a rapid decrease near the phase transition point. An abrupt change seems to occur around point $\hat{M}_2\approx0.85$, which is attributed to numerical inaccuracies. } The behavior of the anomalous Hall conductivity indicates the vanishing of the Weyl nodes, i.e. the phase transitions between the Weyl semimetal phase and the topological trivial phase. We cannot give more details of the phase transitions between the nodal line semimetal to the topological trivial phase through the anomalous Hall conductivity, nevertheless, the calculation of the fermionic spectral functions, the topological invariants and the topological entanglement entropy can achieve this goal and we leave this in a future work\cite{future}.

We then move to the calculation of the free energy. The free energy can be computed through the renormalized action of Eq.\eqref{eq:action}, which is:
\begin{eqnarray}
S_\text{ren}=S_{\text{on-shell}} +S_\text{GH}+S_{c.t.},
\end{eqnarray}
where {$S_{\text{on-shell}}$ is the bulk on-shell action, and $S_\text{GH}$ is the Gibbons-Hawking term
$S_\text{GH}=2\int_{r=r_{\infty}}d^{4}x\sqrt{-\gamma}K$. The counter-term $S_{c.t.}$ is
\begin{eqnarray}
\begin{aligned}
&S_{c.t.}=\int_{r=r_{\infty}}d^{4}x\sqrt{-\gamma}\left(-6-|\Phi_{1}|^{2}-|\Phi_{2}|^{2}+\frac{1}{2}|B_{\mu\nu}|^2+\frac{1}{2}\left(\log r^{2}\right)\left[\frac{1}{4}F^{2}+\frac{1}{4}F_{5}^{2}+\frac{1}{4}\hat{F}^{2}+\right.\right. \\
&~~~\left.\left.+\frac{1}{4}\hat{F}_{5}^{2}+|D_{1\mu}\Phi_{1}|^{2}+|D_{2\mu}\Phi_{2}|^{2}+\left(\frac{1}{3}+\frac{\lambda_{1}}{2}\right)\left(|\Phi_{1}|^{4}+|\Phi_{2}|^{4}\right)+\frac{2}{3}|\Phi_{1}|^{2}|\Phi_{2}|^{2}+|B_{\mu\nu}|^4\right]\right),
\end{aligned}
\end{eqnarray}
where $\gamma_{\mu\nu}$ is the induced metric on the boundary $r=r_{\infty}$, $K$ is the trace of the extrinsic curvature with  $K=\gamma^{\mu\nu}\nabla_\mu n_\nu$ and $n_\nu$ is the outward unit
vector normal to the boundary. The bulk on-shell action is calculated to be a total derivative, which is
\begin{eqnarray}
S_{\text{on-shell}}=\int d^{4}xdr \sqrt{-g}\mathcal{L}=-\int d^{4}x\intop_{0}^{r_{\infty}}dr\left(-\frac{f u h^\prime}{\sqrt{h}}\right)^{\prime}.
\label{free}
\end{eqnarray}
With the field expansion near the UV boundary $r\to \infty$, we obtain the free energy density
\begin{eqnarray}
\frac{\Omega}{V}=-\frac{1}{V}S_{ren}&=&-\frac{9a_{0}^{2}M_{2}^{2}}{8}+\frac{11b^{4}}{9}-\frac{38b^{2}M_{1}^{2}}{9}-\frac{8b^{2}M_{2}^{2}}{9}-\frac{8bb_{xy2}}{3}\nonumber\\
&+&4h_{2}-\frac{7M_{1}^{2}M_{2}^{2}}{18}-\frac{7M_{2}^{4}}{36}+2M_{1}O_{1}+2M_{2}O_{2}-u_{2}.
\label{free2}
\end{eqnarray}
\vspace{0cm}
\begin{figure}[ht!]
  \centering
\includegraphics[width=0.8\textwidth]{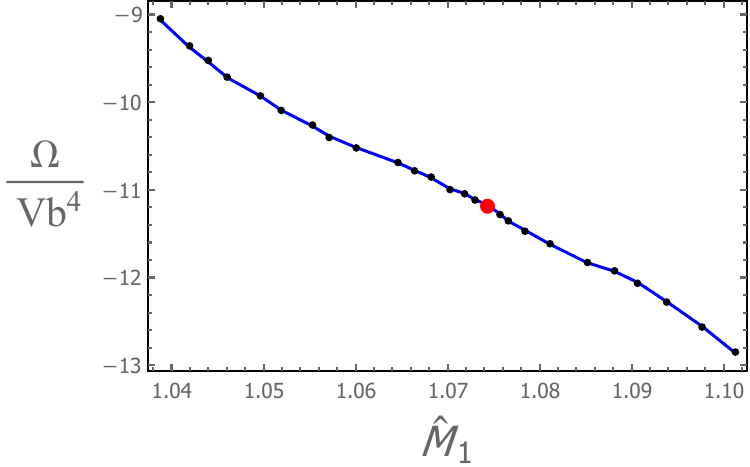}
\vspace{-0.3cm}
  \caption{\small The free energy density as a function of $\hat{M}_1$ at $c/b =1$ and $\hat{M}_2=0.924$ across the critical point. The red dot is the value of free energy density at the double critical red dot in Fig. \ref{fig:02}.}
 \label{fig:04}
\end{figure}

Eq.\eqref{free2} yields the same result as the holographic Weyl semimetal when all parameters corresponding to the nodal line semimetal become zero, i.e. $b=M_1=b_{xy2}=O_1=0$\cite{Landsteiner:2015pdh}. While when $a_0=M_2=O_2=0$ and $h_2=u_2$, the same formula for the free energy can be obtained in the holographic model that only describes a single nodal line semimetal\cite{Liu:2020ymx}. More details can be found in appendix \ref{app:c}.

The topological phase transitions in the holographic Weyl semimetal, nodal line semimetal and the Z$_2$-Weyl semimetals all have been shown to be continuous from the behavior of the free energy. Here as we are having a two dimensional phase diagram, to avoid too much numerics, we will calculate the free energy of this system in a one parameter curve to show the continuity of the phase transition as an example. With a fixed value of $M_2=0.924$, we have shown the free energy density as a function of $M_1$ of the system in  Fig.\ref{fig:04}. As presented in Fig.\ref{fig:02}, when $M_2$ (as well as the value of $c/b=1$) is fixed, the system transitions from the gap-nodal state to the Weyl-gap state as $M_1$ varies. The numerical results of the free energy density indicate that this phase transition process is smooth when crossing the double critical point.

%%%%%%%%%%%%%%%%%%%%%%%%%%
\section{Discussion and Outlook}
\label{sec:4}
%%%%%%%%%%%%%%%%%%%%%%%%%%
We have constructed a holographic model that allows for the coexistence of the Weyl semimetal and the nodal line semimetal, inspired by a weakly coupled effective field theory model. The holographic model allows for the study of the topological system in the strong coupling limit, and the difference in the topological behavior compared to the weak coupling system. 
%comparison of the system's similarity and difference in the weak and strong coupling regimes. 
We have found that topologically nontrivial Weyl and nodal line semimetal states could still coexist in the strong coupling limit and there are nine zero-temperature solutions of the holography model, which correspond to nine different phases of the system. This is qualitatively the same as in the weak coupling field theory, leading to a similar phase diagram. However, the shape of the phase diagram and the quantitative values of the phase boundaries are different in the strong coupling limit. 
We have also calculated the anomalous Hall conductivity, which is one order parameter of the system, which confirms the topological phase transition in the phase diagram. The free energy in the holographic model is also calculated to confirm the continuity of the topological phase transition. 

%, we have found that the topological phase transitions in this system are generally smooth when crossing the phase transition, similar to the weak coupling case. However, the parameter spaces of each phase that enable survival are quite different in the holographic model and the field theory model, particularly in the critical region. 

%\textcolor{purple}{what does the following paragraph mean?}\textcolor{cyan}{discuss}
%The concept of the renormalization group may aid in our comprehension of this. Specifically, in this system, the nodal line semimetal phase is determined by the effective radius of the nodal ring. The Weyl semimetal phase is determined by the distance between the two Weyl nodes, while the gapped phase is determined by the effective mass. In the weak coupling field theory model, the effective radius of the nodal ring, the distance between the two Weyl nodes, and the effective mass are directly affected by the combination of different parameters. Therefore, the phases in the field theory model are highly responsive to changes in parameters. The holographic model relates different phases to distinct geometries on the IR side, while the "effective parameters" are the outcomes on the UV side. The running effects of the solutions cannot be ignored. Therefore, it is worthwhile to examine the compatibility between the field theory model and the holographic model by verifying the renormalization group of the coupling in the weak coupling field theory model.

There are still some open questions that we would like to leave for future work. First, in a similar but four-component weakly coupled field theory model \cite{Ji:2023rua}, the system could have a topological triple degenerate nodal point phase, which has also been observed in several systems in laboratories. It is worth considering whether a holographic model can have this novel topological state \cite{future}. Additionally, the fermionic spectral functions and the topological invariants of the holographic model \eqref{eq:action} need to be calculated to further confirm their topological and transport properties, which we also leave for future work.  Third, could a Weyl Semimetal equipped with multiple pairs of Weyl nodes coexist with the nodal line semimetal, both in the weak couling limit and the strong coupling limit? Finally, it is intriguing to explore the impact of disorder in a system with multiple topological phases.   
%%%%%%%%%%%%%%%%%%%%%%%%%%%%%%
\subsection*{Acknowledgments}
%%%%%%%%%%%%%%%%%%%%%%%%%%%%%%
We thank K. Landsteiner, Y. Liu and X. M. Wu for discussions. This work was supported by the National Natural Science Foundation of China (Grant Nos.12035016).
%%%%%%%%%%%%%%%%%%%%%%%%%%%
%%%%%%%%%%%%%%%%%%%%%%%%%%%
\appendix
\section{The phase of the system}
\label{app:a}
Four free parameters exist in the model \eqref{eq:1Lagrangian}. They are two mass terms $M_1$ and $M_2$, $b_x=b\delta_{\mu}^{x}$ the 1-form field and $b_{xy}=c$ the two-form field. Nine types of spectrums of the system have been shown in Fig.\ref{fig:eight} by different relations between these parameters and we will introduce in details below.

\vspace{.25cm}
\noindent {\bf  $\bullet$ \em The Weyl-nodal phase}
 
$M_1<2c$ and $M_2<b$, a state where the Weyl semimetal and the nodal line semimetal state coexist forms(Fig. \ref{fig:eight}(a)). The location of the two Weyl nodes are $(k_x, k_y, k_z)=\left(\pm\sqrt{b^2-M_{2}^2}\,,0\,,0\right)$ while the radius of the nodal ring equal to $\sqrt{4c^2-M_{1}^2}$.

\vspace{.25cm}
\noindent {\bf $\bullet$ \em The Weyl-critical phases}

Fix $b, c, M_2$ in (a) in Fig.~\ref{fig:eight} and increase $M_1$ the radius of the nodal ring will decrease. When $M_1=2 c$ the nodal ring form a critical Dirac node while the Weyl nodes keep their original form. This phase is shown in (b) in Fig.~\ref{fig:eight}. 

\vspace{.25cm}
\noindent {\bf $\bullet$ \em The Weyl-gap phases}

Increase $M_1$ in (b) of Fig.~\ref{fig:eight}, a topological trivial gap forms as shown in case (c) of Fig.~\ref{fig:eight}.

\vspace{.25cm}
\noindent {\bf $\bullet$ \em The double critical point}

Keep $b, c, M_1$ in (b) of Fig.~\ref{fig:eight} and increase the value of $M_2$, the distance between the two Weyl nodes decrease. When $M_2=b$ the two Weyl nodes merge into one Dirac node. A double critical point has been obtained and shown in (d) in Fig.~\ref{fig:eight}.

\vspace{.25cm}
\noindent {\bf $\bullet$ \em The critical-nodal phases}

Fix $b, c, M_1$ in (a) in Fig.~\ref{fig:eight} and increase $M_2$ leads the two Weyl points annihilate to a critical Dirac point, which is shown in (e) of Fig.~\ref{fig:eight}.

\vspace{.25cm}
\noindent {\bf $\bullet$ \em The critical-gap/gap-critical phases}

Continue to increase $M_2$ or $M_1$ from the double critical point in (d) of Fig.~\ref{fig:eight}, the fourfold-degenerate critical point will be aligned into a pair of gapped bands and one twofold-degenerate critical point. These two cases are shown in the case (f) or (h) in Fig.~\ref{fig:eight}. 

\vspace{.25cm}
\noindent {\bf $\bullet$ \em The gap-nodal phases}

Fix $b, c, M_2$ in Fig. \ref{fig:eight}(e) and increase $M_1$, the Dirac point becomes a gap, which is shown in (g) of Fig.~\ref{fig:eight}.

\vspace{.25cm}
\noindent {\bf $\bullet$ \em The gap-gap phase}

Starting from (f) or (h) in Fig.~\ref{fig:eight}, and increase the corresponding mass parameter, the system would fully gapped, which corresponds to (i) in Fig.~\ref{fig:eight}.

\vspace{0cm}
\begin{figure}[ht!]
  \centering
  \begin{subfigure}[b]{0.33\textwidth}
\includegraphics[width=\textwidth]{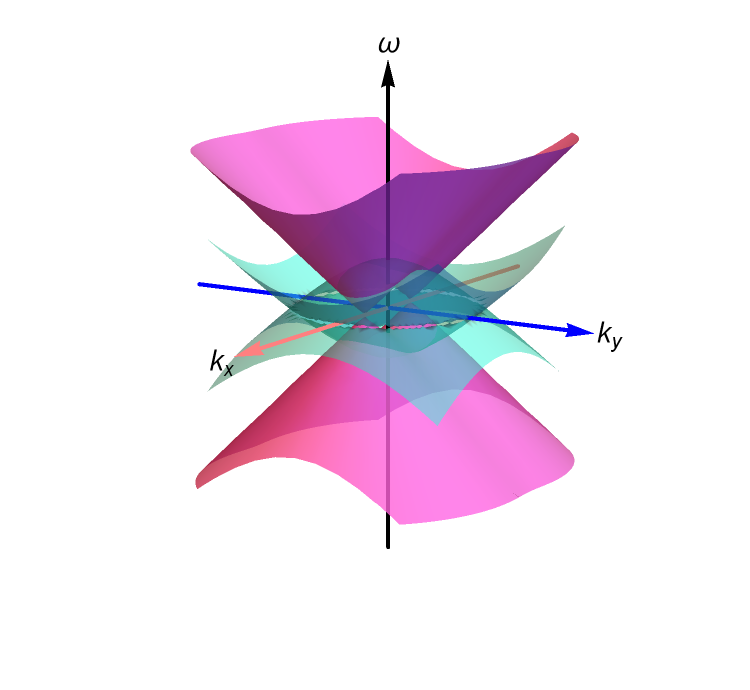}
\caption{\small Weyl-Nodal}
\end{subfigure}
\begin{subfigure}[b]{0.32\textwidth}
\includegraphics[width=\textwidth]{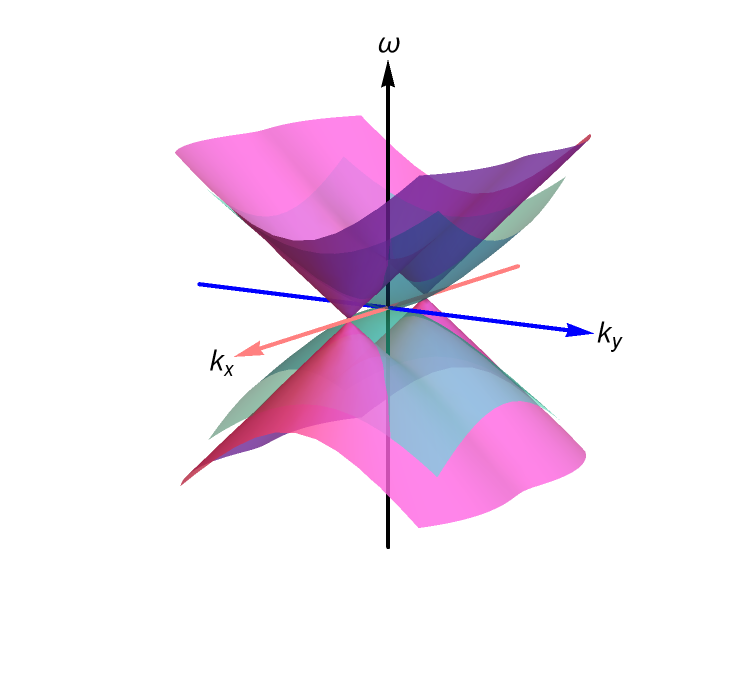}
\caption{\small Weyl-Critical}
\end{subfigure}
\begin{subfigure}[b]{0.33\textwidth}
\includegraphics[width=\textwidth]{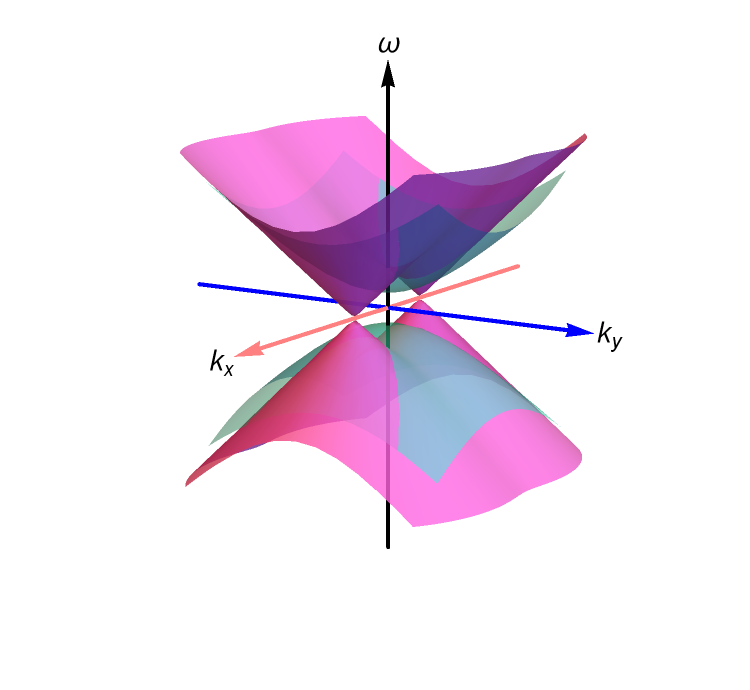}
\caption{\small Weyl-Gap}
\end{subfigure}
\begin{subfigure}[b]{0.33\textwidth}
\includegraphics[width=\textwidth]{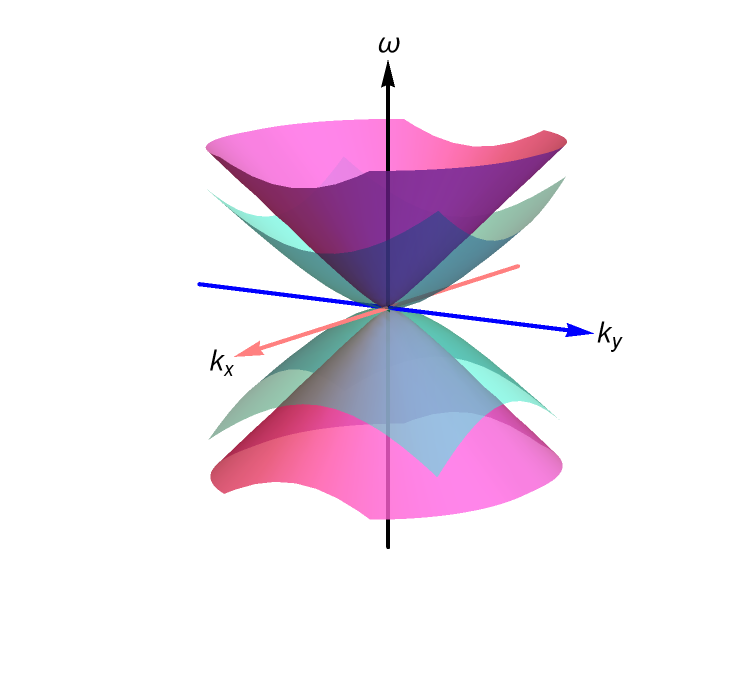}
\caption{\small Dirac Point(Critical)}
\end{subfigure}
\begin{subfigure}[b]{0.32\textwidth}
\includegraphics[width=\textwidth]{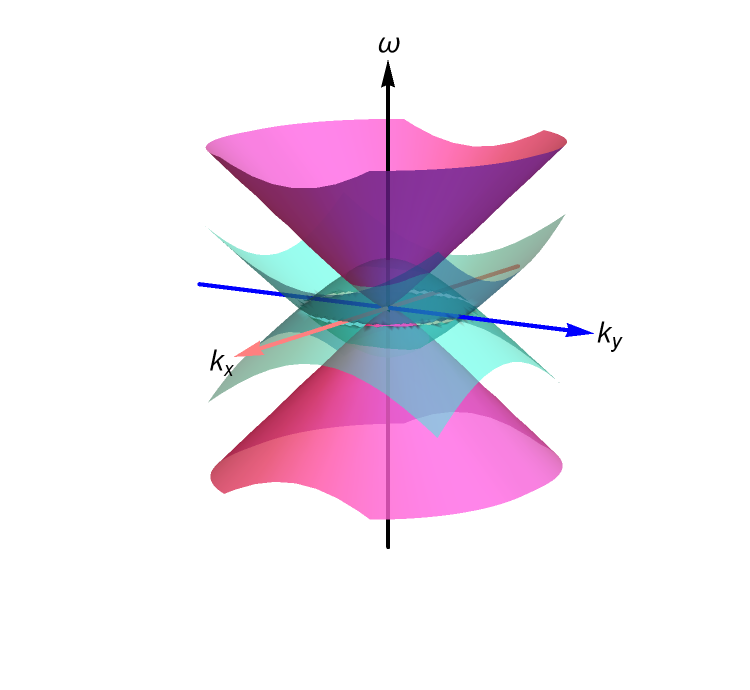}
\caption{\small Critical-Nodal}
 \end{subfigure}
 \begin{subfigure}[b]{0.33\textwidth}
\includegraphics[width=\textwidth]{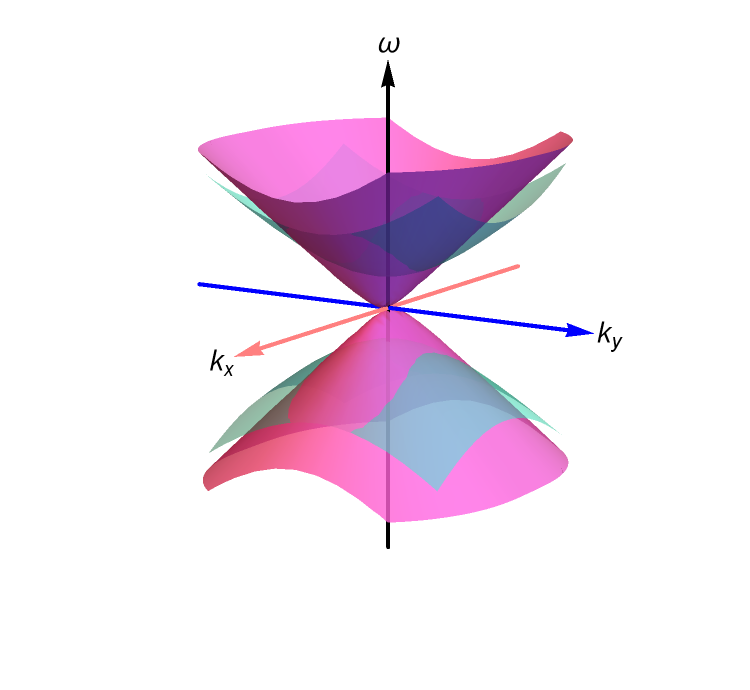}
\caption{\small Critical-Gap}
 \end{subfigure}
 \begin{subfigure}[b]{0.33\textwidth}
\includegraphics[width=\textwidth]{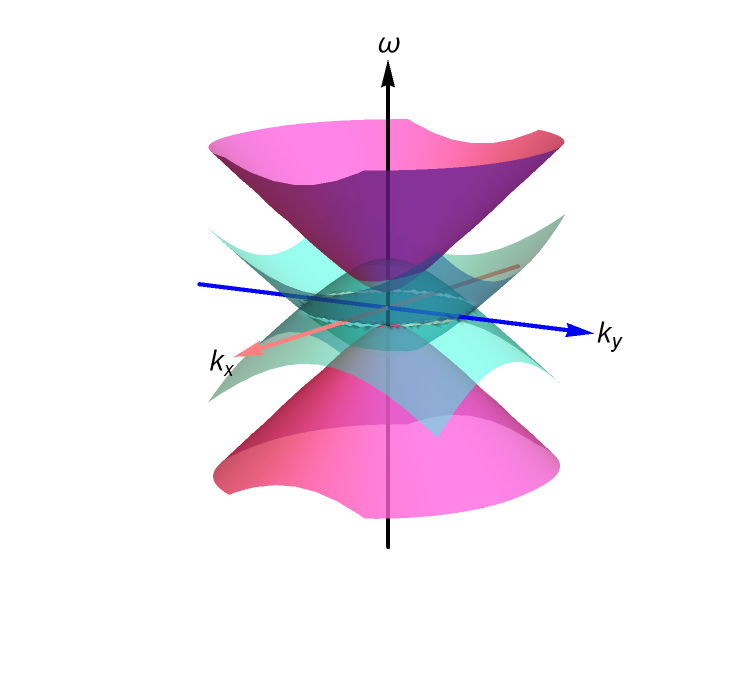}
\caption{\small Gap-Nodal}
 \end{subfigure}
  \begin{subfigure}[b]{0.32\textwidth}
\includegraphics[width=\textwidth]{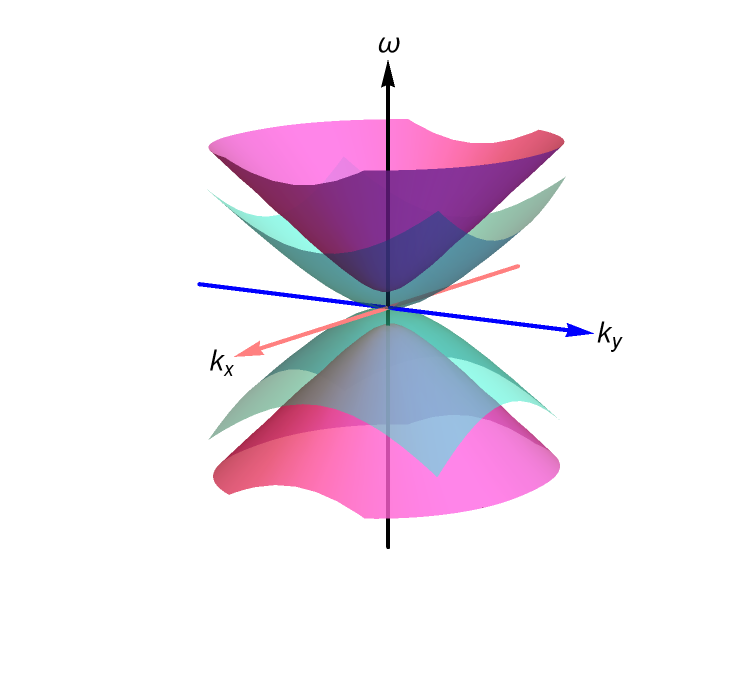}
\caption{\small Gap-Critical}
 \end{subfigure}
  \begin{subfigure}[b]{0.33\textwidth}
\includegraphics[width=\textwidth]{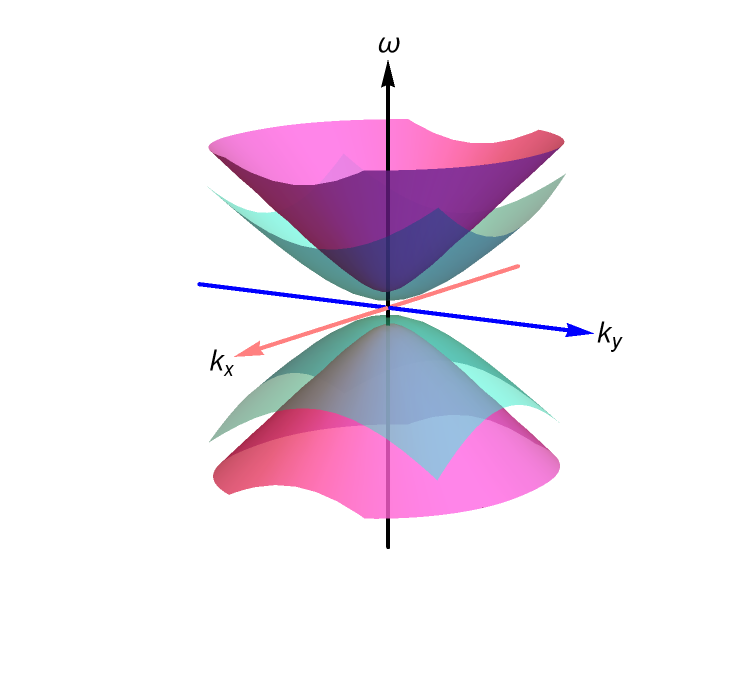}
\caption{\small Gap-Gap}
 \end{subfigure}
 \caption{\small The energy spectrum of \eqref{eq:1Lagrangian} as a function of $k_x$ and $k_y$ with $k_z=0$. From (a) to (i): (a):the system has two Weyl nodes and one nodal line where $M_1<2c$ and $M_2<b$, (b): two Weyl nodes and a critical point where $M_1=2c$ and $M_2<b$, (c): two Weyl nodes and a gap state where $M_1>2c$ and $M_2<b$, (d): a critical Dirac node where $M_1=2c$ and $M_2=b$, (e): a critical Dirac node and one nodal line where $M_1<2c$ and $M_2=b$, (f): a critical Dirac node and a gap state where $M_1>2c$ and $M_2=b$, (g): a gap state and one nodal line where $M_1<2c$ and $M_2>b$, (h): a gap state and one critical point where $M_1=2c$ and $M_2>b$ and (i): fully gap state where $M_1>2c$ and $M_2>b$. }
  \label{fig:eight}
\end{figure}

\section{The equation of motion}
\label{app:b}
%%%%%%%%%%%%%%%%%%%%%%%%%%%%%%%%%%%%%%%%%
The equations of motion from the action (\ref{eq:action}) in the main text are
\bea
\label{eq:ein1}
\begin{aligned}
R_{ab}-\frac{1}{2}g_{ab}(R+12)-T_{ab}&=0\,,\\
\nabla_{b}\mathcal{F}^{ba}+2\alpha\epsilon^{abcde}F_{bc}^{5}\mathcal{F}_{de}+2\beta\epsilon^{abcde}\hat{F}_{bc}^{5}\hat{\mathcal{F}}_{de}&=0\,,\\
\nabla_b F^{ba}_5+\alpha \epsilon^{abcde} (F_{bc}^{5}F_{de}^{5}+\mathcal{F}_{bc}\mathcal{F}_{de}+\mathcal{\hat{F}}_{bc}\mathcal{\hat{F}}_{de}+\frac{1}{3}\hat{F}_{bc}^{5}\hat{F}_{de}^{5})+\frac{\beta}{3} \epsilon^{abcde}\hat{F}_{bc}^{5}\hat{F}_{de}^{5}
&\\
-iq_2\big(\Phi_2^*D^a\Phi_2-(D^a\Phi_2)^*\Phi_2\big)
+\frac{q_3}{\eta}\epsilon^{abcde} B_{bc}B_{de}^*&=0\,,\\
\hat{D}_a \hat{D}^a\Phi_1-\partial_{\Phi_1^*} V_1-\lambda\Phi_1 B_{ab}^*B^{ab}&=0\,,\\
D_a D^a\Phi_2-m_2^2\Phi_2-\lambda_1\Phi_2^{*2}\Phi_2&=0\,,\\
\frac{i}{3\eta} \epsilon_{abcde} H^{cde}-m_3^2 B_{ab}-\lambda \Phi_1^*\Phi_1 B_{ab}&=0\,,
\end{aligned}
\eea
where $T_{ab}$ is the energy-momentum tensor.

with the zero-temperature ansatz\eqref{eq:ansatz}, the equations of motion can be explicitly written as
\bea
\label{eq:background}
u''-\frac{u f''}{f}+\frac{h' u'}{2h}-\frac{u f' h'}{2fh}-\frac{4\lambda\mathcal{B}_{xy}^2 \phi_1^2}{f^2}-\frac{4\lambda\mathcal{B}_{tz}^2 \phi_1^2}{u h}-\frac{4 m_3^2 \mathcal{B}_{xy}^2 }{f^2}-\frac{4 m_3^2\mathcal{B}_{tz}^2}{uh}&=0\, ,\\
-\frac{6}{u}+\frac{m_3^2\mathcal{B}_{tz}^2 }{h u^2}+\frac{m_3^2\mathcal{B}_{xy}^2}{u f^2}+\frac{m_{1}^{2}\phi_{1}^{2}}{2u}+\frac{\lambda\mathcal{B}_{tz}^2 \phi_1^2}{h u^2}+\frac{\lambda\mathcal{B}_{xy}^2 \phi_1^2}{u f^2}+\frac{\lambda_{1}\phi_{1}^{4}+\lambda_{1}\phi_{2}^{4}}{4u}\,\nonumber\\+\frac{1}{2}\left(\phi_{1}^{\prime2}+\phi_{2}^{\prime2}\right)+\frac{m_{2}^{2}\phi_{2}^{2}}{2u}-\frac{q_2^{2}A_{z}^{2}\phi_{2}^{2}}{2h u}-\frac{A_{z}^{\prime2}}{4h}+\frac{f^{\prime}u^{\prime}}{f u}-\frac{f^{\prime2}}{4 f^2}+\frac{f''}{f}+\frac{u''}{2u}&=0\,,\\
-\frac{6}{u}-\frac{m_3^2\mathcal{B}_{tz}^2}{h u^2}+\frac{m_3^2\mathcal{B}_{xy}^2}{u f^2}+\frac{m_{1}^{2}\phi_{1}^{2}}{2u}-\frac{\lambda\mathcal{B}_{tz}^2 \phi_1^2}{h u^2}+\frac{\lambda\mathcal{B}_{xy}^2 \phi_1^2}{u f^2}+\frac{\lambda_{1}\phi_{1}^{4}+\lambda_{1}\phi_{2}^{4}}{4u}\,\nonumber\\-\frac{1}{2}\left(\phi_{1}^{\prime2}+\phi_{2}^{\prime2}\right)+\frac{m_{2}^{2}\phi_{2}^{2}}{2u}+\frac{q_2^{2}A_{z}^{2}\phi_{2}^{2}}{2h u}-\frac{A_{z}^{\prime2}}{4h}-\frac{f^{\prime2}}{4 f^2}+\frac{f^{\prime}u^{\prime}}{2 f u}+\frac{f^{\prime}h^{\prime}}{2f h}+\frac{u^{\prime}h^{\prime}}{4h u}&=0\,,\\
A_{z}^{\prime\prime}+A_{z}^{\prime}\left(\frac{u^{\prime}}{u}-\frac{h^{\prime}}{2h}+\frac{f^{\prime}}{f}\right)-\frac{2q_2^{2}A_{z}\phi_{2}^{2}}{u}&=0\,,\\ -(m_3^2+\lambda \phi_1^2)\mathcal{B}_{xy}+\frac{2 f}{\eta \sqrt{h}}\mathcal{B}_{tz}'&=0\,\\ (m_3^2+\lambda \phi_1^2)\mathcal{B}_{tz}-\frac{2\sqrt{h} u}{\eta f}\mathcal{B}_{xy}'&=0\,\\ \phi_{1}^{\prime\prime}+\phi_{1}^{\prime}\left(\frac{u^{\prime}}{u}+\frac{h^{\prime}}{2h}+\frac{f^{\prime}}{f}\right)-\frac{\phi_{1}}{u}\left(m_1^{2}+\lambda_{1}\phi_{1}^{2}\right)+\frac{2\lambda\mathcal{B}_{tz}^2 \phi_1}{h u^2}-\frac{2\lambda\mathcal{B}_{xy}^2 \phi_1}{u f^2}&=0\,\\ \phi_{2}^{\prime\prime}+\phi_{2}^{\prime}\left(\frac{u^{\prime}}{u}+\frac{h^{\prime}}{2h}+\frac{f^{\prime}}{f}\right)-\frac{\phi_{2}}{u}\left(m_2^{2}+\lambda_{1}\phi_{2}^{2}\right)-\frac{q_2^{2}A_{z}^2\phi_{2}}{h u}&=0. 
\eea

\section{The free energy}
\label{app:c}
With the formula of the free energy of the system (\ref{free}), we need to expand the field at the UV. Close to the UV boundary, we can obtain the following behavior of the fields with $\lambda=1$, $\lambda_1=1/10$, $q_1=1/2$, $q_2=3/2$ and $q_3=1/4$:
\begin{eqnarray}
&&u=r^{2}-\frac{6 b^{2}+M_{1}^{2}+M_{2}^{2}}{3}+\frac{\log r}{180r^{2}}\left(80 b^{4}+23M_{1}^{4}+40 M_{1}^{2}M_{2}^{2}+23M_{2}^{4}\right)+\frac{u_{2}}{r^{2}}+\ldots\nonumber\\
&&f=r^{2}-\frac{M_{1}^{2}+M_{2}^{2}}{3}+\frac{\log r}{180r^{2}}\left(80 b^{4}+23M_{1}^{4}+40M_{1}^{2}M_{2}^{2}+23M_{2}^{4}\right)+\frac{f_{2}}{r^{2}}+\ldots\nonumber\\
&&h=r^{2}-\frac{6 b^{2}+M_{1}^{2}+M_{2}^{2}}{3}+\frac{\log r}{360r^{2}}\left(160 b^{4}+46M_{1}^{4}+80M_{1}^{2}M_{2}^{2}+46M_{2}^{4}+405 a_{0}^2 M_{2}^{2}\right)+\frac{h_{2}}{r^{2}}+\ldots\nonumber\\
&&A_{z}=a_0-\frac{9 M_{2}^{2}a_0^{2}\log r}{4r^{2}}+\frac{a_{2}}{r^{2}}+\ldots\nonumber\\
&&\mathcal{B}_{xy}=br+2b^3\,\frac{\text{ln}(r)}{r}+\frac{b_{xy2}}{r}+\ldots\nonumber\\
&&\mathcal{B}_{tz}=br-2b^3\,\frac{\text{ln}(r)}{r}+\frac{b_{tz2}}{r}+\ldots\nonumber\\
&&\Phi_{1}=\frac{M_{1}}{r}-\frac{\log r}{60r^{3}}\left(23M_{1}^{3}+20M_{1}M_{2}^{2}\right)+\frac{O_{1}}{r^{3}}+\ldots\nonumber\\
&&\Phi_{2}=\frac{M_{2}}{r}-\frac{\log r}{120r^{3}}\left(46M_{2}^{3}+40M_{2}M_{1}^{2}+135M_{2}a_{0}^{2}\right)+\frac{O_{2}}{r^{3}}+\ldots.
\end{eqnarray}
with $f_2=\frac{9a_{0}^{2}M_{2}^{2}}{64}+\frac{7b^{4}}{18}+\frac{5b^{2}M_{1}^{2}}{18}+\frac{b^{2}M_{2}^{2}}{9}+\frac{bb_{xy2}}{3}-\frac{h_{2}}{2}+\frac{7M_{1}^{2}M_{2}^{2}}{36}+\frac{149(M_{1}^{4}+M_{2}^{4})}{1440}-\frac{M_{1}O_{1}}{2}-\frac{M_{2}O_{2}}{2}$ and $\text{\ensuremath{b_{tz2}}}-\frac{1}{6}\left(6b^{3}+7bM_{1}^{2}+bM_{2}^{2}+6\text{\ensuremath{b_{xy2}}}\right)$ from the equation of motion.

Taking into account the boundary terms and performing a Wick rotation, the free energy
density can be obtained as
\begin{eqnarray}
\label{eC2}
\frac{\Omega}{V}&=&\frac{11b^{4}}{9}-\frac{9a_{0}^{2}M_{2}^{2}}{8}-\frac{38b^{2}M_{1}^{2}}{9}-\frac{8b^{2}M_{2}^{2}}{9}-\frac{8bb_{xy2}}{3}\nonumber\\
&+&4h_{2}-\frac{7M_{1}^{2}M_{2}^{2}}{18}-\frac{7M_{2}^{4}}{36}+2M_{1}O_{1}+2M_{2}O_{2}-u_{2}.
\end{eqnarray}

With the calculations of the stress tensor of the dual field theory, we can check the relationship between the total energy density and free energy:
\begin{eqnarray}
T_{\mu\nu}=2(K_{\mu\nu}-\gamma_{\mu\nu}K)+\frac{2}{\sqrt{-\gamma}}\frac{\delta S_\text{c.t.}}{\delta \gamma^{\mu\nu}}\,.
\end{eqnarray}
The total energy density is
\begin{eqnarray}\label{eC4}
\begin{aligned}
\epsilon=\lim_{r\to\infty}\sqrt{-\gamma} \langle T^0_0\rangle&=-\frac{9a_{0}^{2}M_{2}^{2}}{8}+\frac{11b^{4}}{9}-\frac{38b^{2}M_{1}^{2}}{9}-\frac{8b^{2}M_{2}^{2}}{9}-\frac{8bb_{xy2}}{3}\\
&+4h_{2}-\frac{7M_{1}^{2}M_{2}^{2}}{18}-\frac{7M_{2}^{4}}{36}+2M_{1}\phi_{1}+2M_{2}\phi_{2}-u_{2}
\end{aligned}
\end{eqnarray}
Hence, from \eqref{eC2} and \eqref{eC4} we have $\frac{\Omega}{V}=\epsilon$.
%%%%%%%%%%%%%%%%%%%%%%%%%%%

\end{document}